\documentclass[aps]{revtex4}
\usepackage{amsfonts}
\usepackage{amsmath}
\usepackage{amssymb,epsf}
\usepackage{color}

\begin{document}

\title{Thermodynamic instability of nonlinearly charged black holes in
gravity's rainbow}
\author{S. H. Hendi$^{1,2}$\footnote{
email address: hendi@shirazu.ac.ir}, S. Panahiyan$^{1,3}$\footnote{%
email address: sh.panahiyan@gmail.com}, B. Eslam Panah$^{1}$\footnote{%
email address: behzad.eslampanah@gmail.com} and M. Momennia$^{1}$\footnote{%
email address: momennia1988@gmail.com}} \affiliation{$^1$ Physics
Department and Biruni Observatory, College of Sciences, Shiraz
University, Shiraz 71454, Iran\\
$^2$ Research Institute for Astronomy and Astrophysics of Maragha (RIAAM),
Maragha, Iran\\
$^3$ Physics Department, Shahid Beheshti University, Tehran 19839, Iran}

\begin{abstract}
Motivated by the violation of Lorentz invariancy in quantum gravity, we
study black hole solutions in gravity's rainbow in context of Einstein
gravity coupled with various models of nonlinear electrodynamics. We regard
an energy dependent spacetime and obtain related metric functions and
electric fields. We show that there is an essential singularity at the
origin which is covered with an event horizon. We also compute the conserved
and thermodynamical quantities and examine the validity of the first law of
thermodynamics in the presence of rainbow functions. Finally, we investigate
thermal stability conditions for these black hole solutions in context of
canonical ensemble. We show that thermodynamical structure of the solutions
depends on the choices of nonlinearity parameters, charge and energy
functions.
\end{abstract}

\maketitle

\section{Introduction}

One of the interesting dreams of physicists is finding a consistent quantum
theory of gravity. Although there are a lot of attempts to join gravity and
quantum theories together, there is no complete description of the quantum
gravity. On the other hand, it has been shown that the violation of Lorentz
invariancy is an essential primitive rule to construct a quantum theory of
gravity. The Lorentz invariance violation may be expressed in form of
modified dispersion relations \cite%
{Kostelecky,Gambini,Carroll,Amelino,tHooft}. Indeed, regarding various
approaches of quantum gravity, there are evidences which show that the
Lorentz symmetry might be violated in the ultraviolet limit \cite%
{Iengo,Adams,Gripaios,Alfaro,Belich}, and then it only holds in the infrared
limit of quantum theory of gravity. Since the standard energy-momentum
dispersion relation enjoys the Lorentz symmetry, it is expected to modify
this relation in the ultraviolet limit. In fact, it has been observed that
such modification to the standard energy-momentum relation occurs in some
models based on string theory \cite{Kostelecky}, spin-network in loop
quantum gravity (LQG) \cite{Gambini}, spacetime foam \cite{Amelino}, the
discrete spacetime \cite{tHooft}, Horava-Lifshitz gravity \cite%
{HoravaREVD,HoravaLETT}, ghost condensation \cite{Faizal}, non-commutative
geometry \cite{Carroll,Faizal2012}, and double special relativity.

In doubly special relativity, there are two fundamental constants; the
velocity of light and the Planck energy. In this theory, it is not possible
for a particle to attain energy and velocity larger than the Planck energy
and the velocity of light, respectively. The doubly special relativity has
been generalized to curved spacetime, and this doubly general theory of
relativity is called gravity's rainbow \cite{Magueijo1,Magueijo2}. In
gravity's rainbow, the energy of the test particle affects the geometry of
spacetime. It means that gravity has different effects on the particles with
various energies. Hence, the geometry of spacetime is represented by a
family of energy dependent metrics forming a rainbow of metrics. The
gravity's rainbow can be constructed by considering following deformation of
the standard energy-momentum relation
\begin{equation}
E^{2}f^{2}(\varepsilon )-p^{2}g^{2}(\varepsilon )=m^{2},
\end{equation}%
where $\varepsilon =E/E_{P}$ and $E_{P}$ is the Planck energy. The functions
$f(\varepsilon )$ and $g(\varepsilon )$ are called rainbow functions and
they are phenomenologically motivated. The rainbow functions are required to
satisfy
\begin{equation}
\lim_{\varepsilon \rightarrow 0}f(\varepsilon )=1,~\ \ \ \ \ \ \ \
\lim_{\varepsilon \rightarrow 0}g(\varepsilon )=1.
\end{equation}%
where this condition ensures that we have the standard energy-momentum
relation in the infrared limit. It is worthwhile to mention that the
spacetime is probed at an energy $E$, and by definition this can not be
greater than the Planck energy $E_{P}$. It means that if a test particle is
used to probe the geometry of spacetime, then $E$ is the energy of that test
particle, and so $E$ can not become larger than $E_{P}$ \cite{Peng}. It is
worth mentioning that such justification is based on the Standard model of
particle physics. In other words, if a particle is described by standard
model, the upper limit of the Planck energy is enforced and energy functions
will have to satisfy mentioned condition. Whereas, in trans-planckian
physics, such condition could be violated. It means that the particle
probing spacetime could acquire energies larger than the Planck energy. Such
property has been considered and employed in number of papers \cite%
{trans-planckian1,trans-planckian2,trans-planckian3,trans-planckian4,trans-planckian5,trans-planckian6}%
. This consideration requires modifications in structure of the energy
functions as well. However, we will conduct our study with consideration of
the standard model and mentioned conditions for energies that particle could
acquire. Now, it is possible to define an energy dependent deformation of
the metric $\hat{\mathbf{g}}$ as \cite{Peng}
\begin{equation}
\hat{\mathbf{g}}=\eta ^{\mu \nu }e_{\mu }(E)\otimes e_{\nu }(E),
\end{equation}%
where
\begin{equation}
e_{0}(E)=\frac{1}{f(\varepsilon )}\hat{e}_{0},~\ \ \ \ \ \ \ \ e_{i}(E)=%
\frac{1}{g(\varepsilon )}\hat{e}_{i},
\end{equation}%
in which the hatted quantities refer to the energy independent frame.

In recent years, the effects of gravity's rainbow have been investigated in
the context of black hole thermodynamics in literature \cite%
{THrainbow,THrainbow1,THrainbow2,THrainbow3}. The modification in the
thermodynamics of black rings and other black objects in the context of
gravity's rainbow has been investigated in \cite{AliJHEP,AliNUCL}. In
addition, the hydrostatic equilibrium equation in Einstein gravity's rainbow
has been obtained and the maximum mass of neutron stars has been
investigated in Ref. \cite{HendiBEP}. As we consider black holes in
gravity's rainbow, the energy $E$ corresponds to the energy of a quantum
particle in neighborhood of the event horizon, which is emitted in the
Hawking radiation \cite%
{THrainbow,Adler,Cavaglia2003,Cavaglia2004,AmelinoREVD}. On the other hand,
gravity's rainbow holds the usual uncertainty principle \cite{LingLi,LiLing}%
. It is possible to translate the uncertainty principle $\Delta p\geq
1/\Delta x$ into a bound on the energy $E\geq 1/\Delta x$ which $E$ can be
interpreted as the energy of a particle emitted in the Hawking radiation. It
has been shown that the uncertainty in the position of a test particle in
the vicinity of the horizon should be equal to the event horizon radius \cite%
{THrainbow,Adler,Cavaglia2003,Cavaglia2004,AmelinoREVD}
\begin{equation}
E\geq 1/\Delta x\approx 1/r_{+},
\end{equation}
where $E$ is the energy of a particle near the horizon which bounded by the
Planck energy $E_{P}$ and can not increase to arbitrary values. This bound
on the energy modifies temperature and entropy of the black hole in
gravity's rainbow \cite{THrainbow}.

Now, we present various motivations for considering nonlinear
electrodynamics. As we know, most physical systems are inherently nonlinear
in the nature and nonlinear field theories are appropriate tools to
investigate such systems. The main reason to consider the nonlinear
electrodynamics (NED) comes from the fact that these theories are
considerably richer than the Maxwell theory and in special case they reduce
to the linear Maxwell field. In addition, some limitations of the Maxwell
field such as, radiation inside specific materials \cite%
{Lorenci2001,Lorenci2002,Novello2003,Novello2012} and description of the
self-interaction of virtual electron-positron pairs \cite%
{Heisenberg,Yajima,Schwinger}, motivate one to regard NED \cite%
{DelphenichQED,Delphenich}. Also, taking into account the NED, one can
remove both the big bang and black hole singularities \cite%
{Ayon,AyonLETT,Lorenci20022,Dymnikova,Corda2010,Corda2011}. One can find
regular black hole solutions of Einstein gravity in the presence of a
suitable NED \cite{Ayon,AyonLETT,Soleng,Oliveira,Palatnik,Ayon1998}.
Moreover, the effects of NED become indeed very important in superstrongly
magnetized compact objects, such as pulsars and neutron stars \cite%
{Cuesta2004,Cuesta20042,Bialynicka}. In addition, horizonless magnetic
solutions in presence of different nonlinear electromagnetic fields have
been investigated in literature \cite{Magnetic1,Magnetic2}. Besides, an
interesting property which is common to all NED models is the fact that
black object solutions coupled to the NED models enjoy the zeroth and first
laws of thermodynamics.

It is well-known that the electric field of a point-like charge has a
divergency in the origin. To remove this singularity, about eighty years ago
Born and Infeld introduced an interesting kind of NED which is known as
Born-Infeld (BI) nonlinear electrodynamics (BINED) theory \cite%
{Born,BornInfeld}. Then, Hoffmann tried to couple the NED with gravity \cite%
{Hoffmann}. The gravitational fields coupled to BINED have been investigated
for static black holes \cite%
{Dehghani2008,Dehghani2006,Allahverdizadeh,Cai2008,Mazharimousavi,Cai2004},
rotating black objects \cite%
{Hendi2008,Hendi2007,Dehghani2007,Rastegar,Hendi2010}, wormholes \cite%
{Lu,Dehghani2009,Eiroa,Hendi2014}, and superconductors \cite%
{Jing2010,Jing2011,Gangopadhyay2012,Roychowdhury,Gangopadhyay2014,Yao}.
Also, BINED has acquired a new impetus, since it naturally arises in the
low-energy limit of the open string theory \cite%
{Fradkin,Matsaev,Bergshoeff,Callan,Andreev,Leigh}. Recently, two different
BI type models of the NED with logarithmic \cite{Soleng} and exponential
forms \cite{HendiJHEP} have been introduced, which can also remove the
divergency of the electric field near the origin. The logarithmic NED
(LNED), like BI theory, removes divergency of the electric field, while the
exponential NED (ENED) does not cancel the divergency, but its singularity
is much weaker than that in the Maxwell theory. Black object solutions
coupled to LNED and ENED have been studied in literature (for e.g., see \cite%
{Hendi2014,Sheykhi,HendiADV}). Despite of BI type models, another example of
NED is power Maxwell invariant (PMI) field \cite%
{Hassaine2007,Hassaine2008,Hassaine2009,HendiRastegar,HendiLETTB,HendiBE,HendiBERS}%
. The basic motivation of regarding PMI theory comes from the fact that it
is interesting to modify Maxwell theory in such a way that its corresponding
energy-momentum tensor will be conformally invariant. Taking into account
the traceless energy-momentum tensor of electrodynamics, one should regard
conformally invariant Maxwell field which is a subclass of PMI theory.

In this paper, we will study thermal stability of black holes in gravity's
rainbow and see how the presence of rainbow functions modifies stability
conditions and phase transition of black holes. Thermodynamical aspects of
black holes have been among the most interesting subjects since the
pioneering work of Hawking and Beckenstein \cite%
{Hawking,Hawking1,Beckenstein}. The analogy between geometrical properties
of the black holes and thermodynamical variables presents a deep insight
into relations between physical properties of gravity and classical
thermodynamics. Due to this fundamental relation, it is believed that a
consistent theory of quantum gravity could be derived through the use of
thermodynamics of black holes. One of the important subjects of black hole
thermodynamics is thermal stability. In thermal stability the positivity of
heat capacity determines that the black holes are thermally stable. In
addition, the divergency of the heat capacity is a place in which black hole
meets a second order phase transition \cite{stability1,stability2}.
Recently, it was shown that divergencies of the heat capacity also coincide
with phase transitions that are observed in extended phase space \cite%
{extended}.

The outline of the paper is as follow. Next section is devoted to
introducing field equations and their related metric functions. In Sec. \ref%
{Thermo}, conserved and thermodynamic quantities will be obtained and the
validity of the first law of thermodynamics will be examined. Then, the
stability of the solutions and phase transition are investigated through the
canonical ensemble. This paper will be finished with some final remarks.

\section{Field equations and metric function \label{FieldEq}}

The metric describing gravity's rainbow is constructed by considering the
effects of energy of a particle. In other words, using doubly general
relativity and parameterizing spacetime with the ratio of $\varepsilon
=E/E_{p}$, one can construct a rainbow spacetime. Interestingly, this metric
contains specific restriction with regard to mentioned ratio which will be
stated later. Considering mentioned method for building up metric, one can
have the rainbow metric in following form in four dimensions
\begin{equation}
d\tau ^{2}=-ds^{2}=\frac{\Psi \left( r\right) }{f\left( \varepsilon \right)
^{2}}dt^{2}-\frac{1}{g\left( \varepsilon \right) ^{2}}\left( \frac{dr^{2}}{%
\Psi \left( r\right) }+r^{2}d\Omega _{k}^{2}\right) ,  \label{metric}
\end{equation}%
where $d\Omega _{k}^{2}$ represents the line elements of $2$-dimensional
hypersurfaces with the constant curvature $2k$ and volume $V_{2}$ with
following forms
\begin{equation}
d\Omega _{k}^{2}=\left\{
\begin{array}{cc}
d\theta ^{2}+\sin ^{2}\theta d\varphi ^{2} & k=1\text{ } \\
&  \\
d\theta ^{2}+\sinh ^{2}\theta d\varphi ^{2} & k=-1\text{ \ } \\
&  \\
d\theta ^{2}+d\varphi ^{2} & k=0\text{ \ }%
\end{array}%
\right. ,  \label{dOmega}
\end{equation}%
in which $2$-dimensional hypersurface with plane, spherical and hyperbola
symmetries are, respectively, denoted by $k=0$, $k=1$ and $k=-1$.

Our goal is to obtain rainbow solutions in Einstein gravity with
cosmological constant in presence of NED. So the total Lagrangian for this
system is
\begin{equation}
L_{\mathrm{total}}=L_{E}-2\Lambda +L(\mathcal{F}),  \label{Lagrangian}
\end{equation}%
in which the Lagrangian of Einstein gravity is $L_{E}=R$, and $\Lambda $
refers to the cosmological constant. The last term in Eq. (\ref{Lagrangian})
is the Lagrangian of NED, which we consider to be in following forms
\begin{equation}
L(\mathcal{F})=\left\{
\begin{array}{cc}
4\beta ^{2}\left( 1-\sqrt{1+\frac{\mathcal{F}}{2\beta ^{2}}}\right) , &
\text{BINED} \\
&  \\
\beta ^{2}\left[ \exp \left( -\frac{\mathcal{F}}{\beta ^{2}}\right) -1\right]
, & \text{ENED}\vspace{0.1cm} \\
&  \\
-8\beta ^{2}\ln \left( 1+\frac{\mathcal{F}}{8\beta ^{2}}\right) , & \text{%
LNED} \\
&  \\
\left( -\mathcal{F}\right) ^{s}, & \text{PMI}%
\end{array}%
,\right.  \label{L(F)}
\end{equation}%
where $\beta $ and $s$ are nonlinearity parameters, the Maxwell invariant is
$\mathcal{F}=F_{ab}F^{ab}$ in which $F_{ab}=\partial _{a}A_{b}-\partial
_{b}A_{a}$ is the electromagnetic field tensor and $A_{b}$ is the gauge
potential. It is worthwhile to mention that in essence BINED, ENED and LNED
are categorized under a same branch and they are called BI type models of
NED. The series expansion of BI type models for large values of nonlinearity
parameter yields similar results: first term is Maxwell invariant which is
related to Maxwell theory of electromagnetic field, the second term is
quadratic Maxwell invariant coupled with nonlinearity parameter and some
factors which depend on theory under consideration (for explicit forms of
the expansion see Ref. \cite{HendiJHEP}). On the other hand, PMI has
different structure and properties comparing to BI type models. In order to
recover the Maxwell field, one should set $s=1$.

Now, we are in a position to obtain field equations. Applying variational
principle to the Lagrangian (\ref{Lagrangian}), one can find
\begin{equation}
\nabla _{a}\left( \sqrt{-g}L_{\mathcal{F}}F^{ab}\right) =0,  \label{Ftr}
\end{equation}%
\begin{equation}
\Lambda g_{ab}+G_{ab}^{(E)}=\frac{1}{2}g_{ab}L(\mathcal{F})-2L_{\mathcal{F}%
}F_{ac}F_{b}^{c},  \label{FE}
\end{equation}%
where $L_{\mathcal{F}}=\frac{dL(\mathcal{F})}{d\mathcal{F}}$ and $%
G_{ab}^{(E)}=R_{ab}-\frac{1}{2}g_{ab}R$.

Next, due to our interest in electrically charged black holes in gravity's
rainbow, we consider a radial electric field which its related gauge
potential is
\begin{equation}
A_{b}=h\left( r\right) \delta _{b}^{t},
\end{equation}

Using Eqs. (\ref{metric}), (\ref{L(F)}) and (\ref{Ftr}), we obtain following
differential equations
\begin{equation}
\begin{array}{rl}
r\beta ^{2}H^{\prime }-2f(\varepsilon )^{2}g(\varepsilon )^{2}H^{3}+2\beta
^{2}H=0,\;\;\; & \text{BINED} \\
&  \\
r\beta ^{2}H^{\prime }+4rf(\varepsilon )^{2}g(\varepsilon )^{2}H^{2}+2\beta
^{2}H=0,\;\;\; & \text{ENED}\vspace{0.1cm} \\
&  \\
\left( 4r\beta ^{2}+rf(\varepsilon )^{2}g(\varepsilon )^{2}H^{2}\right)
H^{\prime }+8H\beta ^{2}-2f(\varepsilon )^{2}g(\varepsilon
)^{2}H^{3}=0,\;\;\; & \text{LNED} \\
&  \\
2H+\left( 2rs-1\right) H^{\prime }=0,\;\;\; & \text{PMI}%
\end{array}%
,  \label{h(r)}
\end{equation}%
in which $H=H\left( r\right) =\frac{dh\left( r\right) }{dr}$, and prime
denotes derivation with respect to radial coordinate. It is a matter of
calculation to show that
\begin{equation}
H\left( r\right) =\left\{
\begin{array}{cc}
\frac{q}{r^{2}\Gamma }, & \text{BINED} \\
&  \\
\frac{q}{r^{2}}\exp \left( -\frac{1}{2}L_{w}\right) , & \text{ENED}\vspace{%
0.1cm} \\
&  \\
\frac{2\beta ^{2}r^{2}}{qf\left( \varepsilon \right) ^{2}g\left( \varepsilon
\right) ^{2}}\left( \Gamma -1\right) , & \text{LNED} \\
&  \\
\frac{q}{r^{\frac{2}{2s-1}}}, & \text{PMI}%
\end{array}%
,\right.  \label{H(r)}
\end{equation}%
where $L_{w}=LamberW\left( \frac{4q^{2}f(\varepsilon )^{2}g(\varepsilon )^{2}%
}{\beta ^{2}r^{4}}\right)$, $\Gamma =\sqrt{1+\frac{q^{2}f(\varepsilon
)^{2}g(\varepsilon )^{2}}{\beta ^{2}r^{4}}}$ and $q$ is an integration
constant related to the electric charge. In order to have a well-defined
solution with PMI source, we should consider the PMI parameter, $s$ larger
than $1/2$ ($s>1/2$).

By employing Eqs. (\ref{metric}), (\ref{FE}) and (\ref{H(r)}), one can find
metric function for gravity's rainbow in presence of the mentioned NED as
\begin{equation}
\Psi \left( r\right) =k-\frac{m}{r}-\frac{\Lambda r^{2}}{3g(\varepsilon )^{2}%
}+\Upsilon ,  \label{f(r)}
\end{equation}%
with
\begin{equation*}
\Upsilon =\left\{
\begin{array}{cc}
\frac{2\beta ^{2}r^{2}}{3g\left( \varepsilon \right) ^{2}}\left[ 1-\mathfrak{%
F}_{1}\right] , & \text{BINED} \\
&  \\
\frac{\beta ^{2}r^{2}}{6g\left( \varepsilon \right) ^{2}}\left[ \frac{8\sqrt{%
\Gamma ^{2}-1}}{5}\left( L_{w}^{\frac{3}{2}}\mathfrak{F}_{3}+\frac{5\left(
1+L_{w}\right) }{4\sqrt{L_{w}}}\right) -1\right] , & \text{ENED}\vspace{0.1cm%
} \\
&  \\
\frac{4\beta ^{2}r^{2}\left( \Gamma -1\right) }{3g\left( \varepsilon \right)
^{2}}\left[ 5-\frac{\ln \left( \frac{2}{\Gamma +1}\right) }{\left( \Gamma
-1\right) }+4\left( \Gamma -1\right) \mathfrak{F}_{2}\right] , & \text{LNED}
\\
&  \\
-\frac{r^{2}\left( 2s-1\right) ^{2}}{\left( 4s-6\right) g\left( \varepsilon
\right) ^{2}}\left( -\frac{\sqrt{2}\left( 2s-3\right) qf\left( \varepsilon
\right) g\left( \varepsilon \right) }{\left( 2s-1\right) r^{2/\left(
2s-1\right) }}\right) ^{2s}, & \text{PMI}%
\end{array}%
,\right.
\end{equation*}%
where $\mathfrak{F}_{1}={}_{2}F_{1}\left( \left[ \frac{-1}{2},\frac{-3}{4}%
\right] ,\left[ \frac{1}{4}\right] ,1-\Gamma ^{2}\right) $, $\mathfrak{F}%
_{2}={}_{2}F_{1}\left( \left[ \frac{1}{2},\frac{1}{4}\right] ,\left[ \frac{5%
}{4}\right] ,1-\Gamma ^{2}\right) $ and $\mathfrak{F}_{3}={}_{2}F_{1}\left( %
\left[ 1\right] ,\left[ \frac{9}{4}\right] ,\frac{L_{w}}{4}\right) $ are the
hypergeometric functions, and also $m$ is an integration constant related to
total mass of the solutions.

\section{Conserved and Thermodynamic Quantities \label{Thermo}}

Considering obtained solutions for different models of NED, this section is
devoted to calculating conserved and thermodynamical quantities, and study
the effects of gravity's rainbow. Then, we are going to expand our study to
stability of the solutions in canonical ensemble.

Due to the fact that employed metric only contains one temporal killing
vector, one can use the concept of surface gravity for calculating the
temperature on the event horizon ($r_{+}$) which leads to
\begin{equation}
T=\frac{1}{2\pi }\sqrt{\nabla _{\mu }\chi _{\nu }\nabla ^{\mu }\chi ^{\nu }}=%
\frac{1}{4\pi }\frac{g(\varepsilon )}{f(\varepsilon )}\frac{d\Psi (r)}{dr}%
|_{r=r_{+}},  \label{Temp1}
\end{equation}%
where dependency on the rainbow functions indicates that the temperature is
modified. Considering Eqs. (\ref{f(r)}) and (\ref{Temp1}), one can find
\begin{equation}
T=\left\{
\begin{array}{cc}
\begin{array}{c}
\frac{kr_{+}^{2}g\left( \varepsilon \right) ^{2}-2r_{+}^{4}\Lambda +2\beta
^{2}r_{+}^{4}\left( 1-\mathfrak{F}_{1+}\right) -4q^{2}f(\varepsilon
)^{2}g(\varepsilon )^{2}\mathfrak{F}_{2+}}{4\pi f\left( \varepsilon \right)
g\left( \varepsilon \right) r_{+}^{3}},%
\end{array}
& \text{BINED} \\
&  \\
\begin{array}{c}
\frac{\left[ kg\left( \varepsilon \right) ^{2}-r_{+}^{2}\left( \Lambda +%
\frac{\beta ^{2}}{2}\right) \right] +\frac{4\beta qf(\varepsilon
)g(\varepsilon )L_{w+}^{3/2}}{15\left( 1+L_{w+}\right) }\left[ \left(
L_{w+}-5\right) \mathfrak{F}_{3+}-\frac{4}{9}L_{w+}\mathfrak{F}_{4}+\frac{5}{%
4}\left( 1+\frac{3}{L_{w+}^{2}}\right) \right] }{4\pi f\left( \varepsilon
\right) g\left( \varepsilon \right) r_{+}},%
\end{array}
& \text{ENED}\vspace{0.1cm} \\
&  \\
\begin{array}{c}
\frac{\left[ \beta ^{2}\ln \left( \frac{1+\Gamma _{+}}{2}\right) -\Lambda +%
\frac{kg(\varepsilon )^{2}}{r_{+}^{2}}-\frac{\left( \Gamma _{+}^{2}-1\right)
^{2}}{r_{+}}\left( \frac{8\mathfrak{F}_{5}}{45\Gamma _{+}}+\frac{2\beta ^{2}%
}{9\left( \Gamma _{+}^{2}-1\right) }\left[ \frac{\left( 7+5\Gamma
_{+}\right) }{\left( 1-\Gamma _{+}\right) \Gamma _{+}}+2\mathfrak{F}_{2+}%
\right] \right) +\frac{\left( \beta ^{2}r_{+}^{4}\left( 5\Gamma _{+}^{2}-%
\frac{6}{\Gamma _{+}}\right) +1\right) }{3\left( 1-\Gamma _{+}\right)
r_{+}^{5}}\right] r_{+}^{2}}{\pi f(\varepsilon )g(\varepsilon )}%
\end{array}%
, & \text{LNED} \\
&  \\
\frac{\left[ \frac{kg\left( \varepsilon \right) ^{2}}{r_{+}^{2}}-\Lambda -%
\frac{\left( 2s-1\right) }{2}\left( \frac{\sqrt{2}qf\left( \varepsilon
\right) g\left( \varepsilon \right) \left( 2s-3\right) }{\left( 2s-1\right)
r_{+}^{2/\left( 2s-1\right) }}\right) ^{2s}\right] r_{+}}{4\pi f\left(
\varepsilon \right) g\left( \varepsilon \right) }, & \text{PMI}%
\end{array}%
,\right.  \label{Temp2}
\end{equation}%
where $\Gamma _{+}=\Gamma \left\vert _{r=r_{+}}\right. $, $%
L_{w+}=L_{w}\left\vert _{r=r_{+}}\right. $, $\mathfrak{F}_{1+}=\mathfrak{F}%
_{1}\left\vert _{r=r_{+}}\right. $, $\mathfrak{F}_{2+}=\mathfrak{F}%
_{2}\left\vert _{r=r_{+}}\right. $, $\mathfrak{F}_{3+}=\mathfrak{F}%
_{3}\left\vert _{r=r_{+}}\right. $, $\mathfrak{F}_{4}={}_{2}F_{1}\left( %
\left[ 2\right] ,\left[ \frac{13}{4}\right] ,\frac{L_{w+}}{4}\right) $ and $%
\mathfrak{F}_{5}={}_{2}F_{1}\left( \left[ \frac{3}{2},\frac{5}{4}\right] ,%
\left[ \frac{9}{4}\right] ,1-\Gamma _{+}^{2}\right) $.

In order to study entropy of the solutions, one can use the area law. It is
easy to show that the entropy is
\begin{equation}
S=\frac{r_{+}^{2}}{4g(\varepsilon )^{2}}.  \label{entropy}
\end{equation}

On the other hand, even with the modifications in metric, the entropy is
independent of the electromagnetic fields. It is notable that although there
is no trace of electromagnetic field in explicit form of the entropy,
horizon radius is affected by electromagnetic field under consideration.

As for total charge of the solutions, one can employ the Gauss law.
Considering this approach, one can show that for BI type models, results are
the same. In other words, in case of BI type models, the total electric
charge is independent of nonlinearity parameter. Whereas for the PMI case,
the total charge is modified and depends on PMI parameter
\begin{equation}
Q=\left\{
\begin{array}{cc}
\frac{f\left( \varepsilon \right) }{4\pi g\left( \varepsilon \right) }q, &
\text{BI type models} \\
&  \\
\frac{s\left( 2s-1\right) \left( \frac{\sqrt{2}\left( 2s-3\right) f\left(
\varepsilon \right) g\left( \varepsilon \right) }{\left( 2s-1\right) }%
\right) ^{2s}q^{2s-1}}{8\left( 3-2s\right) \pi f\left( \varepsilon \right)
g\left( \varepsilon \right) ^{3}}, & \text{PMI}%
\end{array}%
\right. .  \label{Q}
\end{equation}

In order to obtain electric potential, we can calculate it on the horizon
with respect to a reference
\begin{equation}
U=A_{\mu }\chi ^{\mu }\left\vert _{r\rightarrow \infty }\right. -A_{\mu
}\chi ^{\mu }\left\vert _{r\rightarrow r_{+}}\right. ,  \label{U1}
\end{equation}%
which leads to
\begin{equation}
U=\left\{
\begin{array}{cc}
\frac{q}{r_{+}}\mathfrak{F}_{2+}, & \text{BINED} \\
&  \\
\frac{4\beta \sqrt{L_{w+}}\left[ \frac{45\left( 3+L_{w+}\right) }{8}+\frac{%
9L_{w+}(4+L_{w+})\mathfrak{F}_{3+}}{4}+L_{w+}^{2}\mathfrak{F}_{4}\right]
r_{+}}{135g(\varepsilon )f(\varepsilon )(1+L_{w})}, & \text{ENED}\vspace{%
0.1cm} \\
&  \\
\frac{8\beta \left( \Gamma _{+}-1\right) \left[ 10\mathfrak{F}_{2+}+\left(
\Gamma _{+}^{2}-1\right) \mathfrak{F}_{5}\right] r_{+}}{45q}+\frac{2\beta %
\left[ 6\left( 1-\Gamma _{+}\right) +\left( 8-5\Gamma _{+}\right) \right]
r_{+}}{9qf\Gamma _{+}}, & \text{LNED} \\
&  \\
qr_{+}^{\frac{^{2s-3}}{2s-1}}, & \text{PMI}%
\end{array}%
\right. .  \label{U}
\end{equation}

It is worthwhile to mention that for $s\geq \frac{3}{2}$, the gauge
potential is not well-behaved, asymptotically. Therefore, we have both upper
and lower limits for the nonlinearity parameter of PMI theory ($\frac{1}{2}%
<s<\frac{3}{2}$).

It is straightforward to show that the total finite mass of this black hole
is
\begin{equation}
M=\frac{m}{8\pi f(\varepsilon )g(\varepsilon )}.  \label{M}
\end{equation}

Here, we give more details for examination of the first law of
thermodynamics. Evaluating the metric function on the event horizon ($\psi
(r=r_{+})=0$), one can obtain geometrical mass ($m$) as a function of $r_{+}$
and $q$. Inserting $m(r_{+},q)$ in Eq. (\ref{M}), one finds
\begin{equation}
M=\left\{
\begin{array}{cc}
\frac{\left[ \frac{3k}{2}g(\varepsilon )^{2}+r_{+}^{2}\left( \beta ^{2}\left[
1-\mathfrak{F}_{1+}\right] -\frac{\Lambda }{2}\right) \right] r_{+}}{12\pi
f(\varepsilon )g(\varepsilon )^{3}}, & \text{BINED} \\
&  \\
\frac{\left[ kg(\varepsilon )^{2}-r_{+}^{2}\left( \Lambda +\frac{\beta ^{2}}{%
2}\right) +\frac{\beta qf(\varepsilon )g(\varepsilon )}{3\sqrt{L_{w+}}}%
\left( \left( 1+L_{w+}\right) +\frac{4}{5}L_{w+}^{2}\mathfrak{F}_{3+}\right) %
\right] r_{+}}{8\pi f(\varepsilon )g(\varepsilon )^{3}}, & \text{ENED}%
\vspace{0.1cm} \\
&  \\
\frac{\frac{9kr_{+}^{2}g(\varepsilon )^{2}}{4}+\beta ^{2}r_{+}^{4}\left[
4\left( \Gamma _{+}^{2}-1\right) \mathfrak{F}_{2+}-5\Gamma _{+}+3\ln
(1+\Gamma _{+})\right] -3\left[ \frac{\Lambda }{4}+\beta ^{2}\left( \ln 2-%
\frac{5}{3}\right) \right] }{18\pi r_{+}f(\varepsilon )g(\varepsilon )^{3}},
& \text{LNED} \\
&  \\
\frac{\left[ 2\left( kg(\varepsilon )^{2}-\frac{\Lambda r_{+}^{2}}{3}\right)
-\frac{\left( 2s-1\right) ^{2}}{\left( 2s-3\right) }\left( \frac{-\sqrt{2}%
\left( 2s-3\right) qf(\varepsilon )g(\varepsilon )}{\left( 2s-1\right)
r_{+}^{2/\left( 2s-1\right) }}\right) ^{2s}r_{+}^{2}\right] r_{+}}{16\pi
f(\varepsilon )g(\varepsilon )^{3}}, & \text{PMI}%
\end{array}%
\right. .
\end{equation}

Now, we use Eqs. (\ref{entropy}) and (\ref{Q}) to obtain $M=M(S,Q)$ with the
following forms
\begin{equation}
M=\left\{
\begin{array}{cc}
\frac{2\left[ \frac{3k}{8}+S\left( \beta ^{2}-\frac{\Lambda }{2}\right)
-S\beta ^{2}\mathfrak{F}_{1\Delta }\right] \sqrt{S}}{3\pi f(\varepsilon )},
& \text{BINED} \\
&  \\
\frac{4\left[ \pi \beta Q\mathcal{L}_{w}^{3/2}\mathfrak{F}_{3\mathcal{L}%
}+\left( \frac{15k}{16}-\frac{5S}{8}\left( 2\Lambda +\beta ^{2}\right)
\right) +\frac{5\pi \beta Q\left( 1+\mathcal{L}_{w}\right) }{4\sqrt{\mathcal{%
L}_{w}}}\right] \sqrt{S}}{15\pi f(\varepsilon )}, & \text{ENED}\vspace{0.1cm}
\\
&  \\
\frac{-16\pi ^{2}Q^{2}\mathfrak{F}_{2\Delta }+\frac{3S\left( 4S\Lambda
-3k\right) }{4}+12S^{2}\beta ^{2}\left[ \ln \left( \frac{2}{1+\Delta }%
\right) +\frac{5\left( \Delta -1\right) }{3}\right] }{9\pi f(\varepsilon )%
\sqrt{S}}, & \text{LNED} \\
&  \\
\frac{\left[ \frac{3k}{4}-S\Lambda -\frac{3S\left( -2\right) ^{s}\left(
2s-1\right) ^{2}}{2\left( 2s-3\right) }\left( \frac{-\pi Qf(\varepsilon
)^{s\left( 3-2s\right) }}{s2^{s-2}S}\right) ^{2s/(2s-1)}\right] \sqrt{S}}{%
3\pi f(\varepsilon )}, & \text{PMI}%
\end{array}%
\right. ,
\end{equation}%
where $\mathfrak{F}_{1\Delta }=\mathfrak{F}_{1}\left\vert _{\Gamma =\Delta
}\right. $,$\mathfrak{F}_{2\Delta }=\mathfrak{F}_{2}\left\vert _{\Gamma
=\Delta }\right. $ and $\mathfrak{F}_{3\mathcal{L}}=\mathfrak{F}%
_{3}\left\vert _{L_{w}=\mathcal{L}_{w}}\right. $. Also $\Delta $ and $L_{w}$
are in the following forms
\begin{eqnarray*}
\Delta &=&\sqrt{1+\frac{\pi ^{2}Q^{2}}{\beta ^{2}S^{2}}}, \\
\mathcal{L}_{w} &=&LamberW\left( \frac{4\pi ^{2}Q^{2}}{\beta ^{2}S^{2}}%
\right) .
\end{eqnarray*}

Now, we are in a position to study the validity of the first law of
thermodynamics. Here, we should calculate $\left( \frac{\partial M}{\partial
S}\right) _{Q}$ and $\left( \frac{\partial M}{\partial Q}\right) _{S}$, and
then use Eqs. (\ref{entropy}) and (\ref{Q}) to convert them as functions of $%
r_{+}$ and $q$. After some simplifications, one finds that these quantities
are, respectively, the same as temperature and potential which were obtained
in Eqs. (\ref{Temp2}) and (\ref{U}). Hence, although rainbow functions
affected thermodynamic and conserved quantities, the first law remains valid
as
\begin{equation}
dM=\left( \frac{\partial M}{\partial S}\right) _{Q}dS+\left( \frac{\partial M%
}{\partial Q}\right) _{S}dQ.  \label{1st law}
\end{equation}

\section{Thermodynamic stability}

In this section, we investigate thermal stability conditions of nonlinearly
charged black hole solutions in gravity's rainbow. To do so, we investigate
the heat capacity which is known as canonical ensemble. Thermal stability
conditions are indicated by the sign of heat capacity. The positivity of
heat capacity guarantees thermally stable solutions, whereas the negative
heat capacity is denoted as an unstable state. Another advantage of studying
the heat capacity is investigation of the phase transition point. There is a
limitation for horizon radius ($r_{+0}$) as well as a phase transition that
are obtainable through calculating the root and divergence point of the heat
capacity.

One can use following relation for calculating the heat capacity
\begin{equation}
C_{Q}=T\left( \frac{\partial S}{\partial T}\right) _{Q}=T\frac{\left( \frac{%
\partial S}{\partial r_{+}}\right) _{Q}}{\left( \frac{\partial T}{\partial
r_{+}}\right) _{Q}}.  \label{CQ}
\end{equation}

Using obtained values for temperature and entropy for various models of NED
(Eqs. (\ref{Temp2}) and (\ref{entropy})), one can find following relations
\begin{equation}
C_{Q}=\left\{
\begin{array}{cc}
\frac{\left[ \mathfrak{F}_{1+}+2\left( \Gamma _{+}^{2}-1\right) \mathfrak{F}%
_{2_{+}}-\frac{\left( kg\left( \varepsilon \right) ^{2}+\left( 2\beta
^{2}-\Lambda \right) r_{+}^{2}\right) }{2\beta ^{2}r_{+}^{2}}\right] }{4\pi
\beta ^{2}g\left( \varepsilon \right) ^{2}\left[ \mathfrak{F}_{1+}+\frac{%
4\left( \Gamma _{+}^{2}-1\right) ^{2}}{5}\mathfrak{F}_{5}+\frac{\left(
kg\left( \varepsilon \right) ^{2}-\left( 2\beta ^{2}-\Lambda \right)
r_{+}^{2}\right) }{2\beta ^{2}r_{+}^{2}}\right] r_{+}^{2}}, & \text{BINED}
\\
&  \\
\frac{r_{+}^{2}\left( 1+L_{w_{+}}\right) ^{2}\mathcal{A}_{1}}{4\pi
g^{2}(\varepsilon )\mathcal{A}_{2}}, & \text{ENED} \\
&  \\
\frac{\beta ^{2}\Gamma _{+}^{5}(1-\Gamma _{+})\left( \Gamma
_{+}^{2}-1\right) \mathcal{B}_{1}r_{+}^{2}}{128\pi g^{2}(\varepsilon )\left(
\mathcal{B}_{2}+\mathcal{B}_{3}\right) }, & \text{LNED} \\
&  \\
\frac{\left( \left( s-\frac{1}{2}\right) \left[ \frac{-\sqrt{2}%
qf(\varepsilon )g(\varepsilon )\left( 2s-3\right) }{\left( 2s-1\right)
r_{+}^{2/\left( 2s-1\right) }}\right] ^{2s}-\Lambda +\frac{kg(\varepsilon
)^{2}}{r_{+}^{2}}\right) r_{+}^{2}}{4\pi g(\varepsilon )^{2}\left( \left( s+%
\frac{1}{2}\right) \left[ \frac{-\sqrt{2}qf(\varepsilon )g(\varepsilon
)\left( 2s-3\right) }{\left( 2s-1\right) r_{+}^{2/\left( 2s-1\right) }}%
\right] ^{2s}-\Lambda -\frac{kg(\varepsilon )^{2}}{r_{+}^{2}}\right) }, &
\text{PMI}%
\end{array}%
\right. ,  \label{CQ1}
\end{equation}%
where $A_{1}$, $A_{2}$, $B_{1}$, $B_{2}$ and $B_{3}$ are
\begin{eqnarray*}
\mathcal{A}_{1} &=&\beta qf(\varepsilon )g(\varepsilon )L_{w+}^{2}\left[
\left( 5-L_{w_{+}}\right) \mathfrak{F}_{3+}+\frac{4L_{w_{+}}\mathfrak{F}_{4}%
}{9}-\frac{5\left( 3+L_{w+}^{2}\right) }{4L_{w+}^{2}}\right] \\
&&-\frac{15L_{w+}^{1/2}\left( 1+L_{w+}\right) \left[ kg^{2}(\varepsilon
)-r_{+}^{2}\left( 2\Lambda +\beta ^{2}\right) \right] }{4},
\end{eqnarray*}%
\begin{eqnarray*}
\mathcal{A}_{2} &=&-60\beta qf(\varepsilon )g(\varepsilon )L_{w+}^{2}\left\{
\left( L_{w+}^{3}+3L_{w+}^{2}-9L_{w+}-35\right) \mathfrak{F}_{5}-\frac{16}{3}%
L_{w+}\left( L_{w+}+\frac{4}{3}\right) \mathfrak{F}_{4}\right. \\
&&\left. -\frac{32}{117}L_{w+}^{2}\left( L_{w+}+1\right) \mathfrak{F}_{6}+%
\frac{5\left( L_{w+}-1\right) }{4L_{w+}^{2}}\left(
L_{w+}^{3}+5L_{w+}^{2}+15L_{w+}+3\right) \right\} \\
&&+\frac{15L_{w+}^{1/2}\left[ kg^{2}(\varepsilon )+r_{+}^{2}\left( 2\Lambda
+\beta ^{2}\right) \right] }{8}\left(
L_{w+}^{3}+3L_{w+}^{2}+3L_{w+}+1\right) ,
\end{eqnarray*}%
\begin{eqnarray*}
\mathcal{B}_{1} &=&9\ln \left( 2\left( 1+\Gamma _{+}\right) \right) +4\left(
1-\Gamma _{+}^{2}\right) \left[ \mathfrak{F}_{2+}+\frac{2}{5}\left( 1-\Gamma
_{+}^{2}\right) \mathfrak{F}_{5}\right] -\frac{9}{4\beta ^{2}}\left( \Lambda
-\frac{kg(\varepsilon )^{2}}{r_{+}^{2}}\right) \\
&&+\frac{3\left( 6-\Gamma _{+}-5\Gamma _{+}^{3}\right) }{\left( 1+\Gamma
_{+}\right) \Gamma _{+}}+\frac{2\left( 1+\Gamma _{+}\right) \left( 7+5\Gamma
_{+}\right) }{\Gamma _{+}},
\end{eqnarray*}%
\begin{eqnarray*}
\mathcal{B}_{2} &=&\frac{9\Gamma _{+}^{3}}{16\beta ^{6}r_{+}^{4}}\left[
\frac{\beta ^{2}\left( 3-\Gamma _{+}^{2}\right) }{2}-\left( \Gamma
_{+}^{2}-\Gamma _{+}-1\right) \right] -\frac{\ln \left( 1+\Gamma _{+}\right)
\Gamma _{+}^{2}}{3r_{+}^{20}} \\
&&-\frac{\beta ^{6}\Gamma _{+}^{5}\left( \Gamma _{+}^{2}-1\right) ^{2}}{2}%
\left\{ \frac{\left( \frac{\left( 1+5\Gamma _{+}^{2}\right) }{4}\mathfrak{F}%
_{2+}-\frac{\left( \Gamma _{+}^{2}-1\right) \left( 3+5\Gamma _{+}^{2}\right)
}{5}\mathfrak{F}_{5}\right) }{\left( \Gamma _{+}^{2}-1\right) }+\frac{\left(
\Gamma _{+}^{2}-1\right) \left( 1+\Gamma _{+}^{2}\right) \mathfrak{F}_{7}}{%
3\beta ^{2}g(\varepsilon )^{4}}\right. \\
&&-\frac{9}{16\beta ^{4}\Gamma _{+}^{2}}\left[ \frac{\left( \frac{17\beta
^{2}}{3}-\frac{\Gamma _{+}^{2}}{r_{+}^{4}}-\frac{3\left( 3-\Gamma
_{+}^{2}\right) \ln 2}{4\beta ^{2}r_{+}^{6}}\right) }{\left( \Gamma
_{+}^{2}-1\right) ^{2}}+\frac{\left( \Lambda +\frac{kg(\varepsilon )^{2}}{%
r_{+}^{2}}+\frac{32\beta ^{2}\left( \Gamma _{+}^{2}-1\right) }{3}\right) }{%
2\left( \Gamma _{+}^{2}-1\right) ^{2}}\right. \\
&&\left. \left. +\frac{3\left( \frac{kg(\varepsilon )^{2}}{r_{+}^{2}}-\frac{%
\Lambda }{3}\right) }{4}+\frac{\left( \Lambda +\frac{10kg(\varepsilon )^{2}}{%
3r_{+}^{2}}+\frac{320\beta ^{2}\left( \Gamma _{+}^{2}-1\right) }{9}\right) }{%
4\left( \Gamma _{+}^{2}-1\right) }\right] \right\} ,
\end{eqnarray*}%
\begin{eqnarray*}
\mathcal{B}_{3} &=&\frac{\beta ^{4}\Gamma _{+}^{2}}{3}\left\{ \frac{%
-45\Gamma _{+}^{5}}{32}-\frac{3\left( \Gamma _{+}^{2}-1\right) \Gamma _{+}}{%
10}\left[ -\frac{5\left( \Gamma _{+}^{2}-1\right) \Gamma _{+}^{2}}{2}\left(
\mathfrak{F}_{2+}+\frac{2\left( 1-\Gamma _{+}^{2}\right) \mathfrak{F}_{5}}{5}%
\right) \right. \right. \\
&&\left. +\frac{45}{32\beta ^{2}}\left( 4\left[ 3-\beta ^{2}\Gamma
_{+}^{2}\ln 2+\frac{kg(\varepsilon )^{2}}{4r_{+}^{2}}+\frac{23\beta
^{2}\left( \Gamma _{+}^{2}-1\right) }{9}\right] +\frac{\left( \Gamma
_{+}^{2}-1\right) \left[ kg(\varepsilon )^{2}-\Lambda r_{+}^{2}\right] }{%
r_{+}^{2}}\right) \right] \\
&&+\left( \Gamma _{+}^{2}-1\right) \Gamma _{+}^{4}\left( \frac{9\mathfrak{F}%
_{2+}}{4}-\frac{12\left( \Gamma _{+}^{2}-1\right) \mathfrak{F}_{5}}{5}%
+\left( \Gamma _{+}^{2}-1\right) ^{2}\mathfrak{F}_{7}\right) \\
&&-\frac{27\left( \Gamma _{+}^{2}-1\right) ^{2}}{64\beta ^{2}}\left[ \frac{%
4\left( 2+\beta ^{4}\Gamma _{+}^{4}\ln 2\right) }{\beta ^{2}\left( \Gamma
_{+}^{2}-1\right) ^{2}}+\frac{\left( \Lambda +\frac{kg(\varepsilon )^{2}}{%
r_{+}^{2}}+\frac{38\beta ^{2}\left( \Gamma _{+}^{2}-1\right) }{3}\right) }{%
\left( \Gamma _{+}^{2}-1\right) ^{2}}\right. \\
&&\left. \left. +\frac{2\left( \Lambda +\frac{kg(\varepsilon )^{2}}{r_{+}^{2}%
}+\frac{34\beta ^{2}\left( \Gamma _{+}^{2}-1\right) }{9}\right) }{\left(
\Gamma _{+}^{2}-1\right) }+\frac{\left( \Lambda +\frac{kg(\varepsilon )^{2}}{%
r_{+}^{2}}+\frac{10\beta ^{2}\left( \Gamma _{+}^{2}-1\right) }{9}\right) }{%
q^{2}}\right] \right\} ,
\end{eqnarray*}%
in which $F_{7}={}_{2}F_{1}\left( \left[ \frac{5}{2},\frac{9}{4}\right] ,%
\left[ \frac{13}{4}\right] ,1-\Gamma _{+}^{2}\right) $.

In order to investigate thermal stability of the solutions, we may regard
explicit functional forms of the rainbow functions $f(\varepsilon)$ and $%
g(\varepsilon)$. The choices of these functions are motivated from various
theoretical and phenomenological considerations. Here, we refer to more
important forms which are based on interesting phenomenology.

The first model is related to constant speed of light and one may use it to
solve the horizon problem \cite{Magueijo1,Magueijo2}. The functional form of
both rainbow functions are the same
\begin{equation}
f(\varepsilon )=g(\varepsilon )=\frac{1}{1-\lambda \varepsilon }.
\label{Mod4}
\end{equation}

In addition, motivated by the results of loop quantum gravity and also
non-commutative geometry, the rainbow functions are given by \cite%
{Amelino2013,Jacob2010}
\begin{equation}
f(\varepsilon )=1,\quad g(\varepsilon )=\sqrt{1-\eta \varepsilon ^{n}}.
\label{Mod1}
\end{equation}

Also, it was shown that in non-commutative geometry context, it is better to
regard a Gaussian trial functional form (exponential function) for $%
f(\varepsilon )$ to avoid a regularization/renormalization scheme \cite%
{trans-planckian6,Garattini2011}. On the other hand, based on the hard
spectra from gamma-ray burster's, one may consider the rainbow functions
\cite{Amelino} with the following forms
\begin{equation}
f(\varepsilon )=\frac{e^{\xi \varepsilon }-1}{\xi \varepsilon },\text{ \ \ }%
g(\varepsilon )=1.  \label{loop}
\end{equation}

\begin{figure}[tbp]
$%
\begin{array}{cc}
\epsfxsize=7cm \epsffile{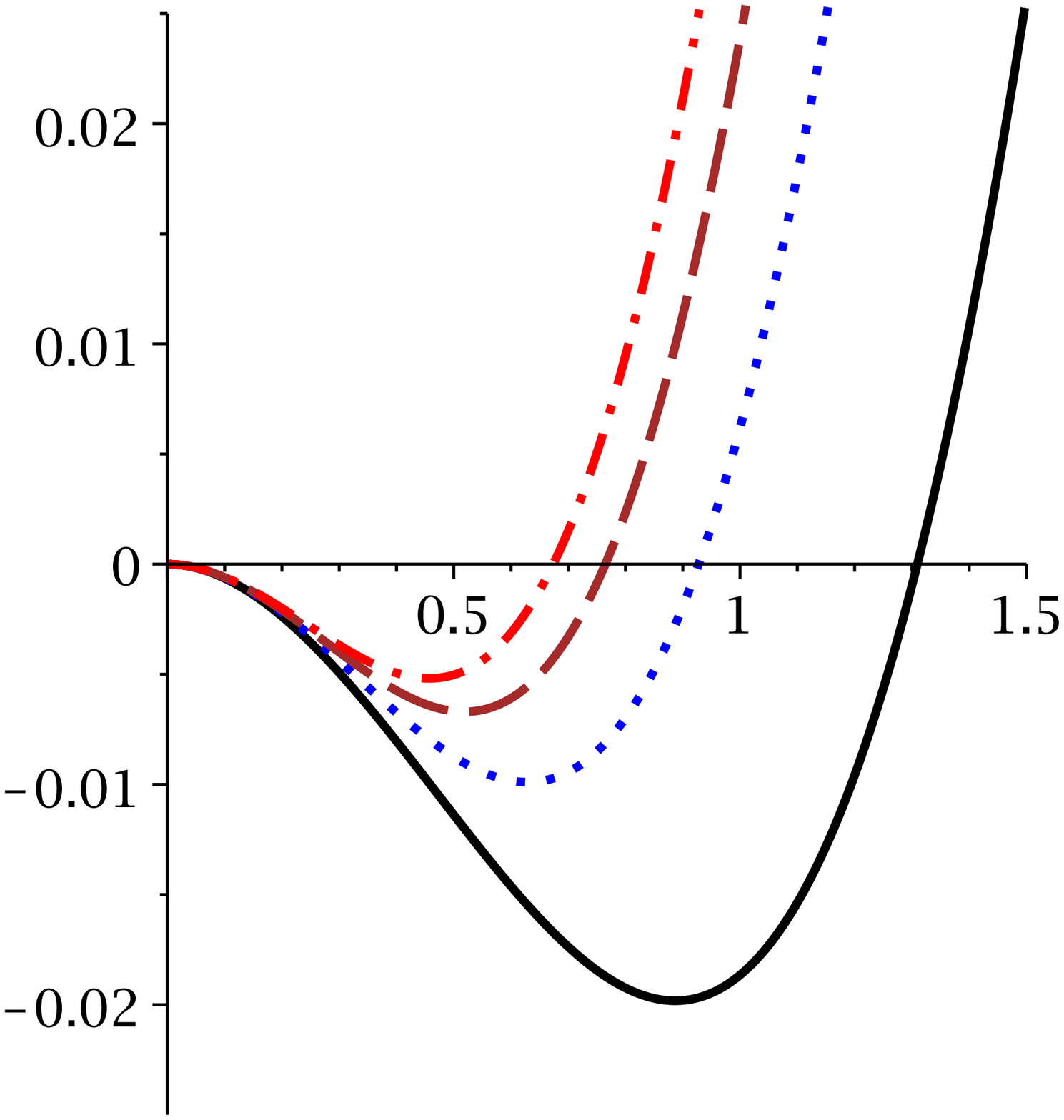} & \epsfxsize=7cm %
\epsffile{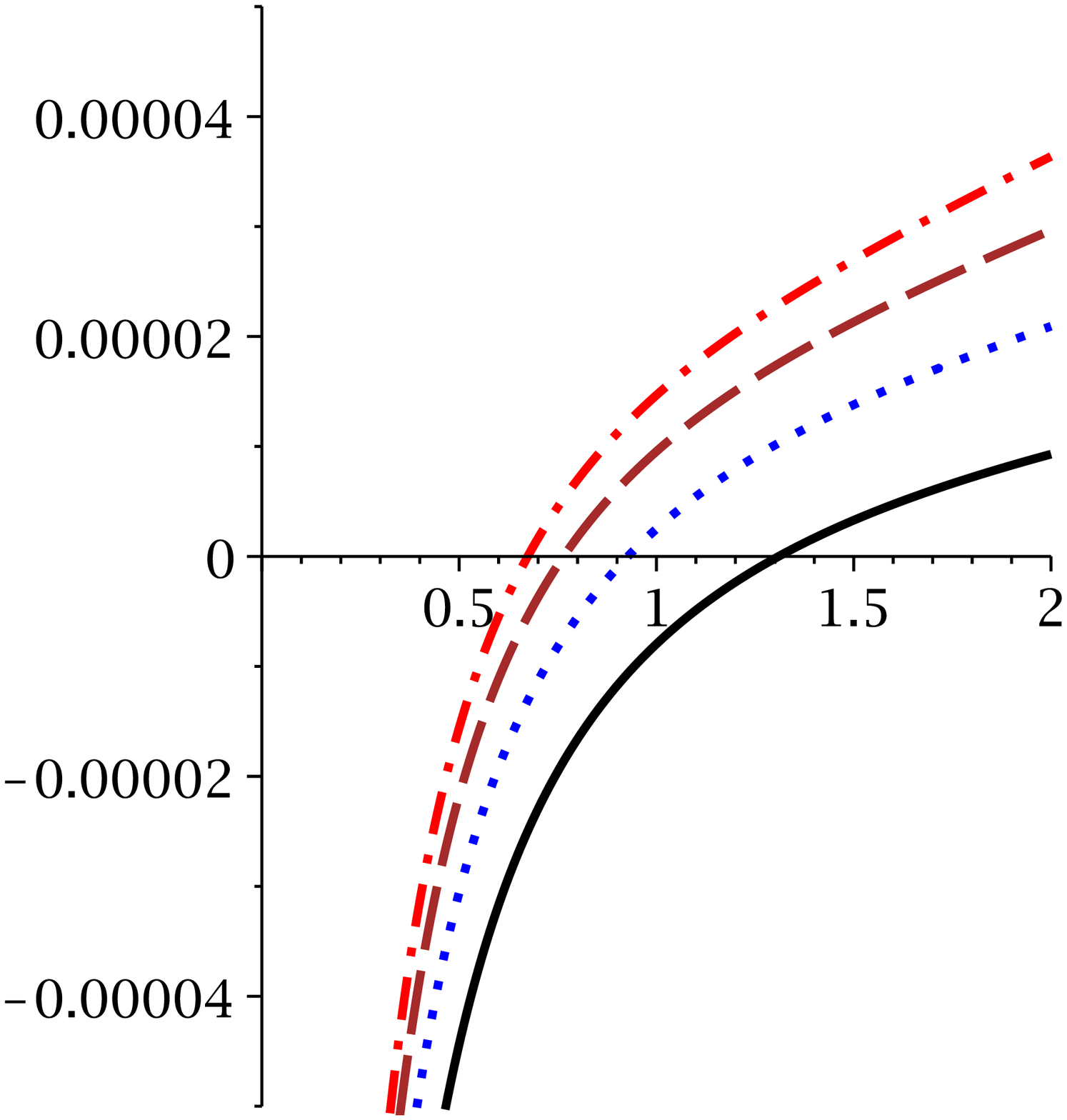}%
\end{array}
$%
\caption{\textbf{ENED branch:} $C_{Q}$ (left panel) and $T$ (right panel)
versus $r_{+}$ for $k=1$, $l=1$, $\protect\beta=2$, $\protect\varepsilon=0.2$
and $q=1$.\newline
$\protect\xi=0.5$ (continues line), $\protect\xi=1$ (dotted line), $\protect%
\xi=1.5$ (dashed line) and $\protect\xi =2$ (dotted-dashed line).}
\label{Fig1}
\end{figure}

\begin{figure}[tbp]
$%
\begin{array}{cc}
\epsfxsize=7cm \epsffile{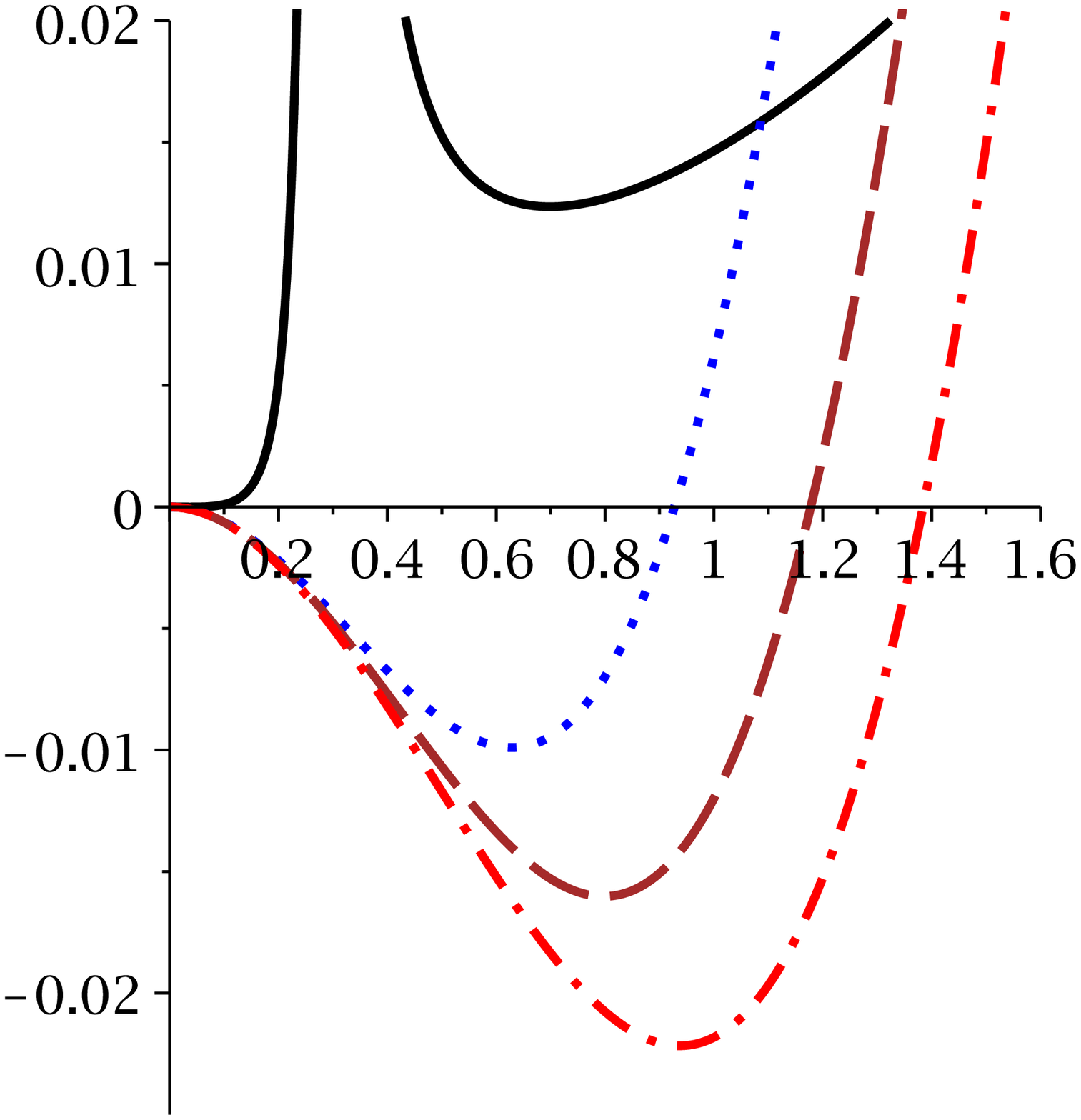} & \epsfxsize=7cm %
\epsffile{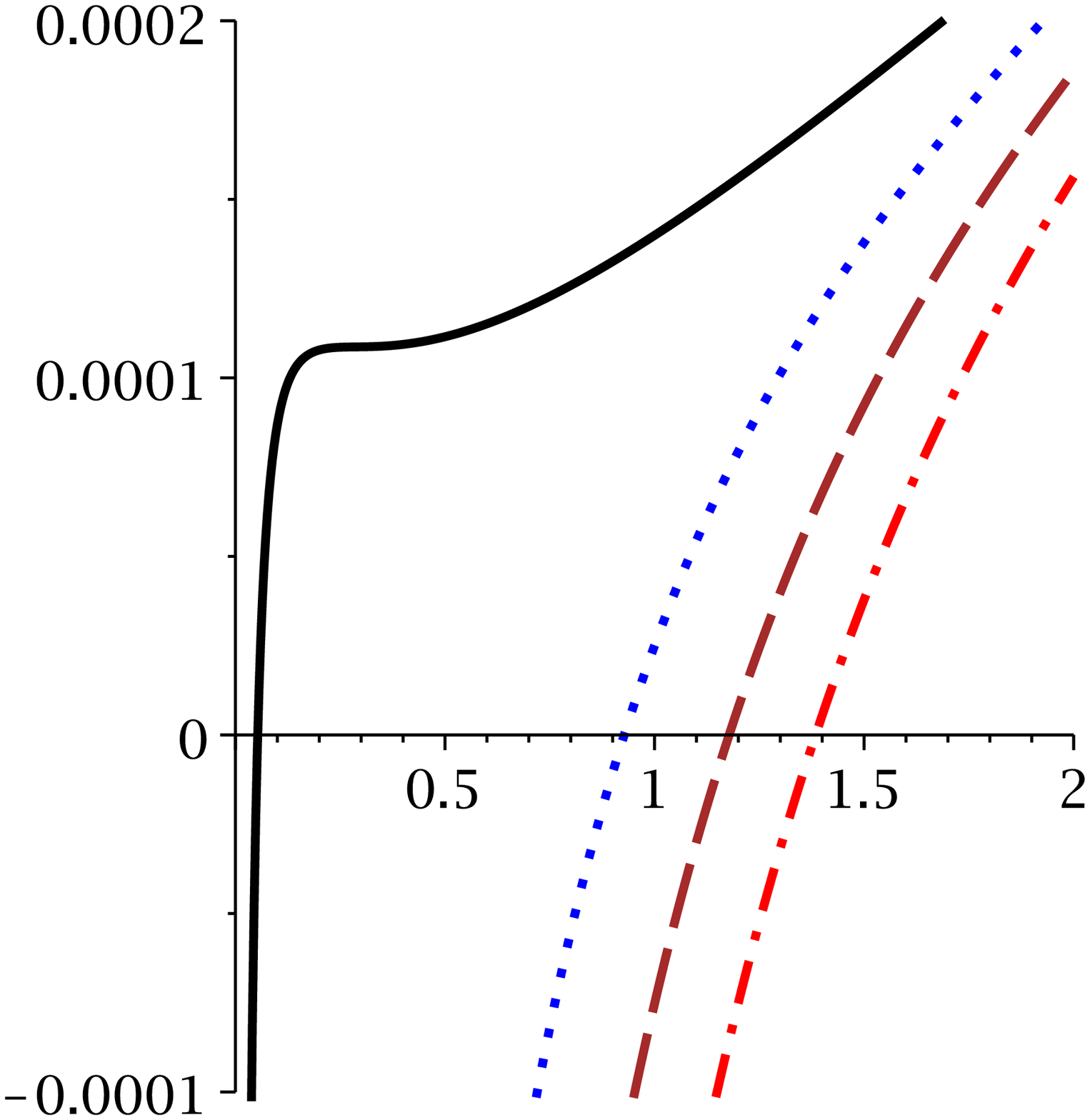}%
\end{array}
$%
\caption{\textbf{ENED branch:} $C_{Q}$ (left panel) and $T$ (right panel)
versus $r_{+}$ for $k=1$, $l=1$, $\protect\beta=2$, $\protect\varepsilon=0.2$
and $\protect\xi=1$.\newline
$q=0.1$ (continues line), $q=1$ (dotted line), $q=1.5$ (dashed line) and $%
q=2 $ (dotted-dashed line).}
\label{Fig2}
\end{figure}

\begin{figure}[tbp]
$%
\begin{array}{ccc}
\epsfxsize=5.5cm \epsffile{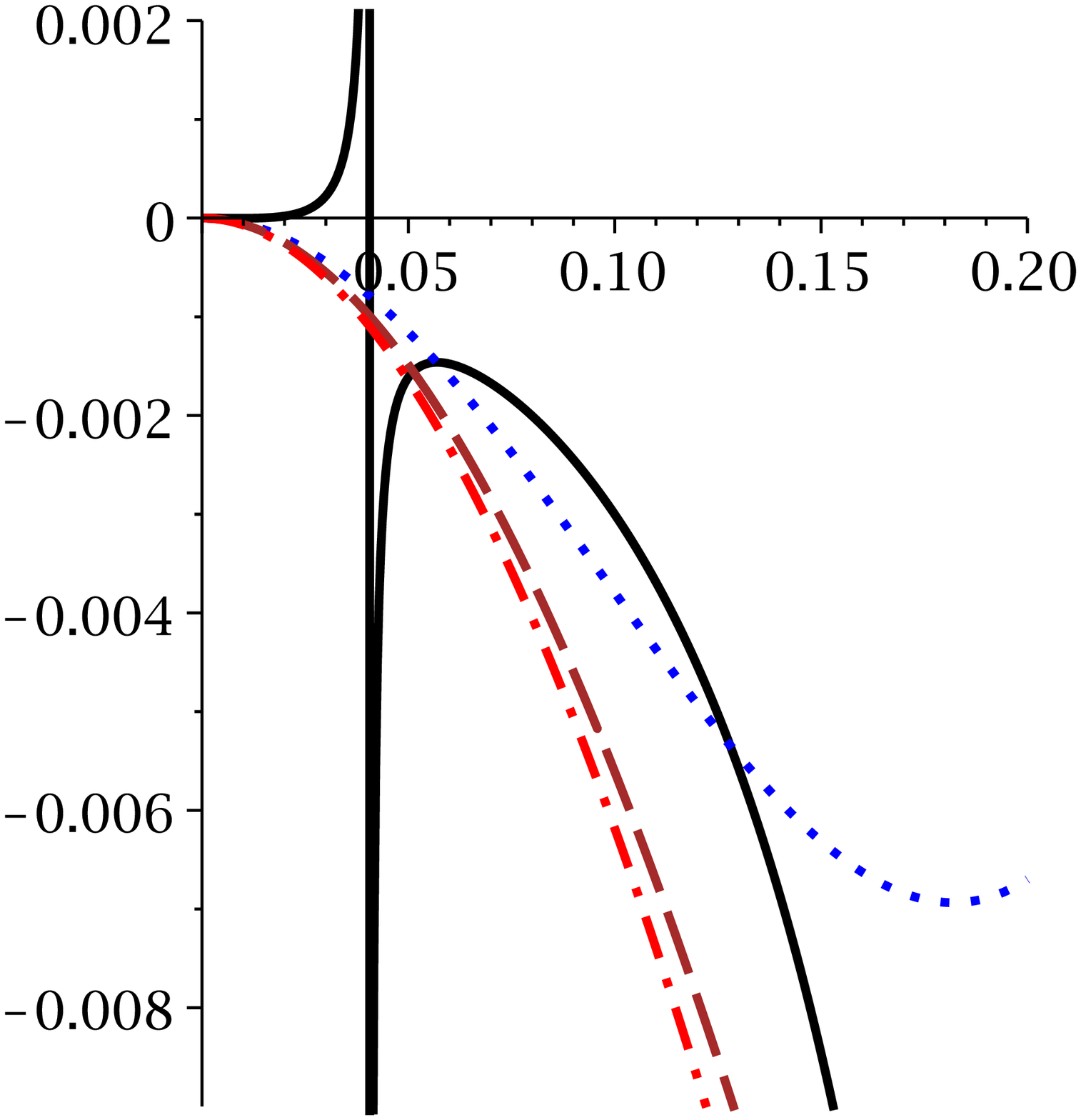} & \epsfxsize=5.5cm %
\epsffile{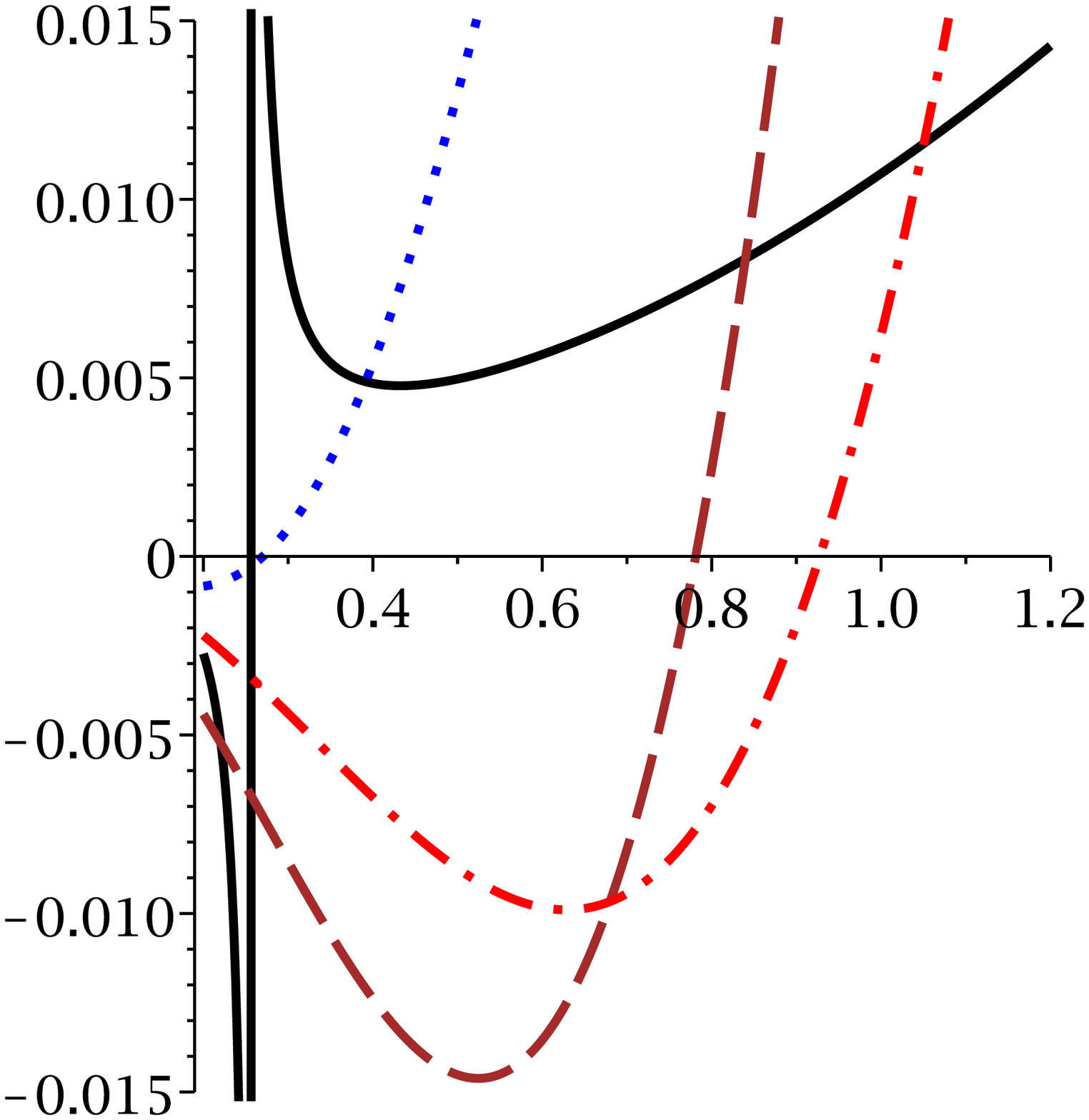} & \epsfxsize=5.5cm %
\epsffile{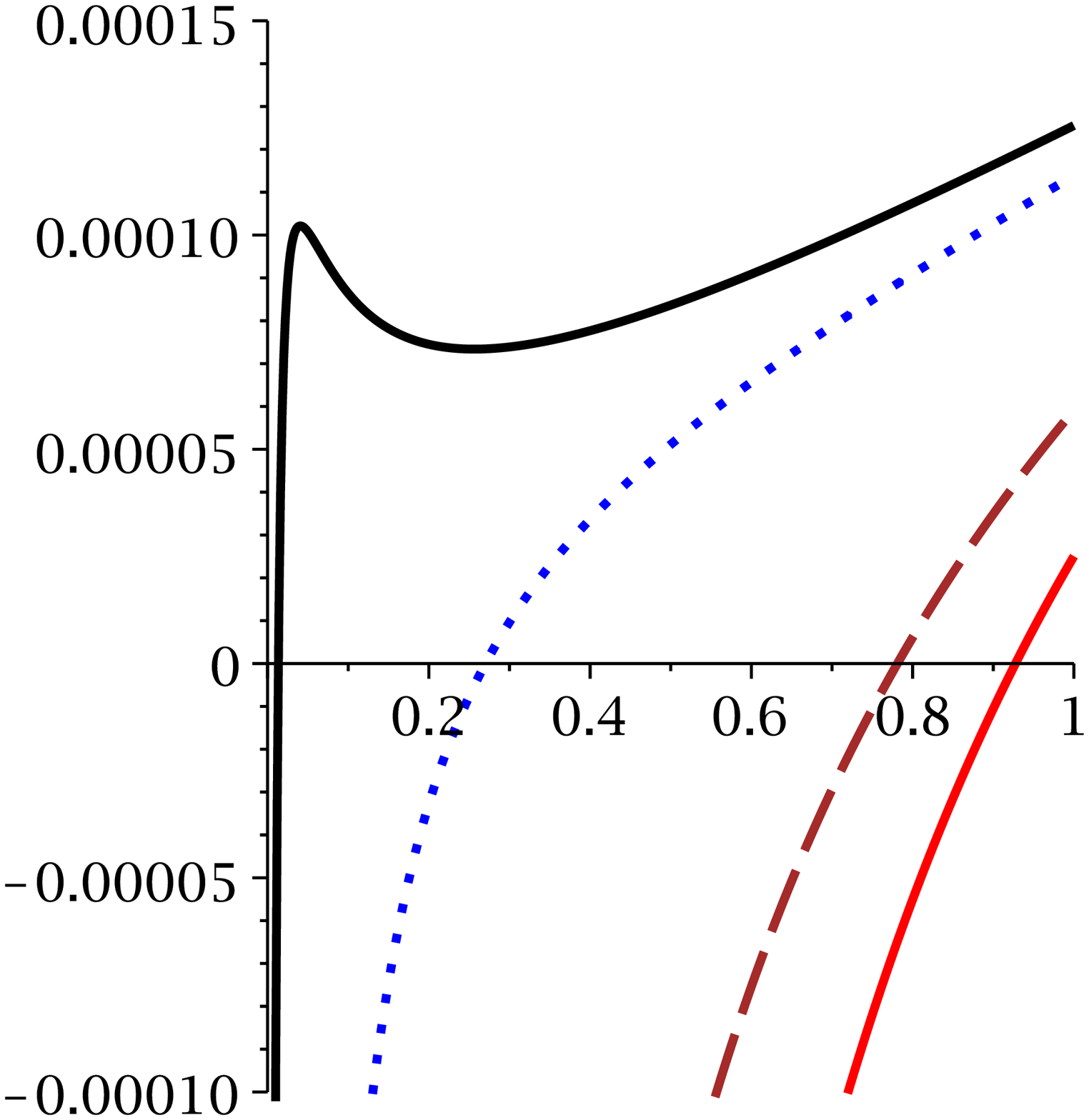}%
\end{array}
$%
\caption{\textbf{ENED branch:} $C_{Q}$ (left and middle panels) and $T$
(right panel) versus $r_{+} $ for $k=1$, $l=1$, $q=1$, $\protect\varepsilon %
=0.2$ and $\protect\xi=1$ "for different scales".\newline
$\protect\beta=0.1$ (continues line), $\protect\beta=0.2$ (dotted line), $%
\protect\beta=1$ (dashed line) and $\protect\beta=2$ (dotted-dashed line).}
\label{Fig3}
\end{figure}

\begin{figure}[tbp]
$%
\begin{array}{cc}
\epsfxsize=5.5cm \epsffile{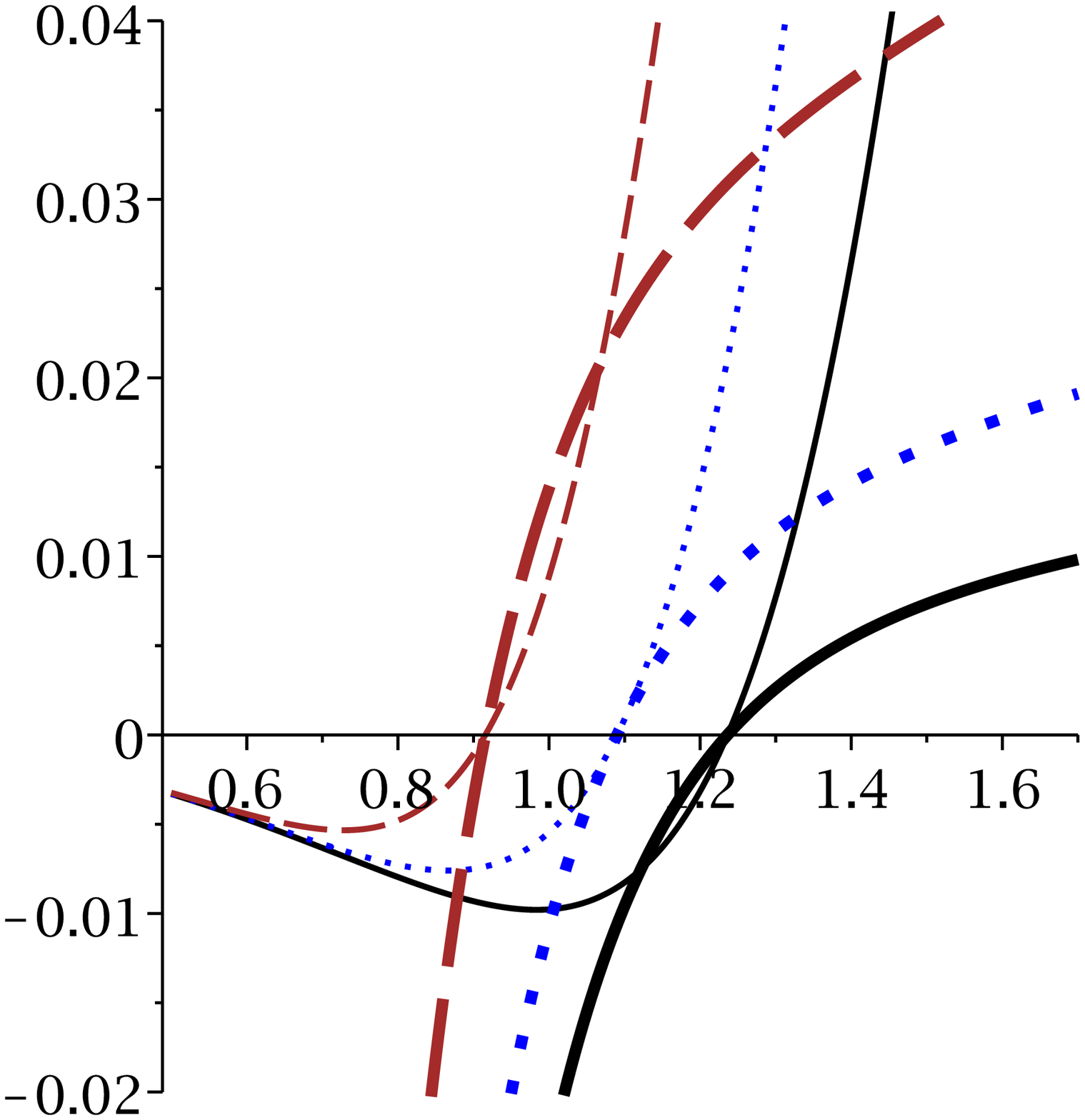} & \epsfxsize=5.5cm %
\epsffile{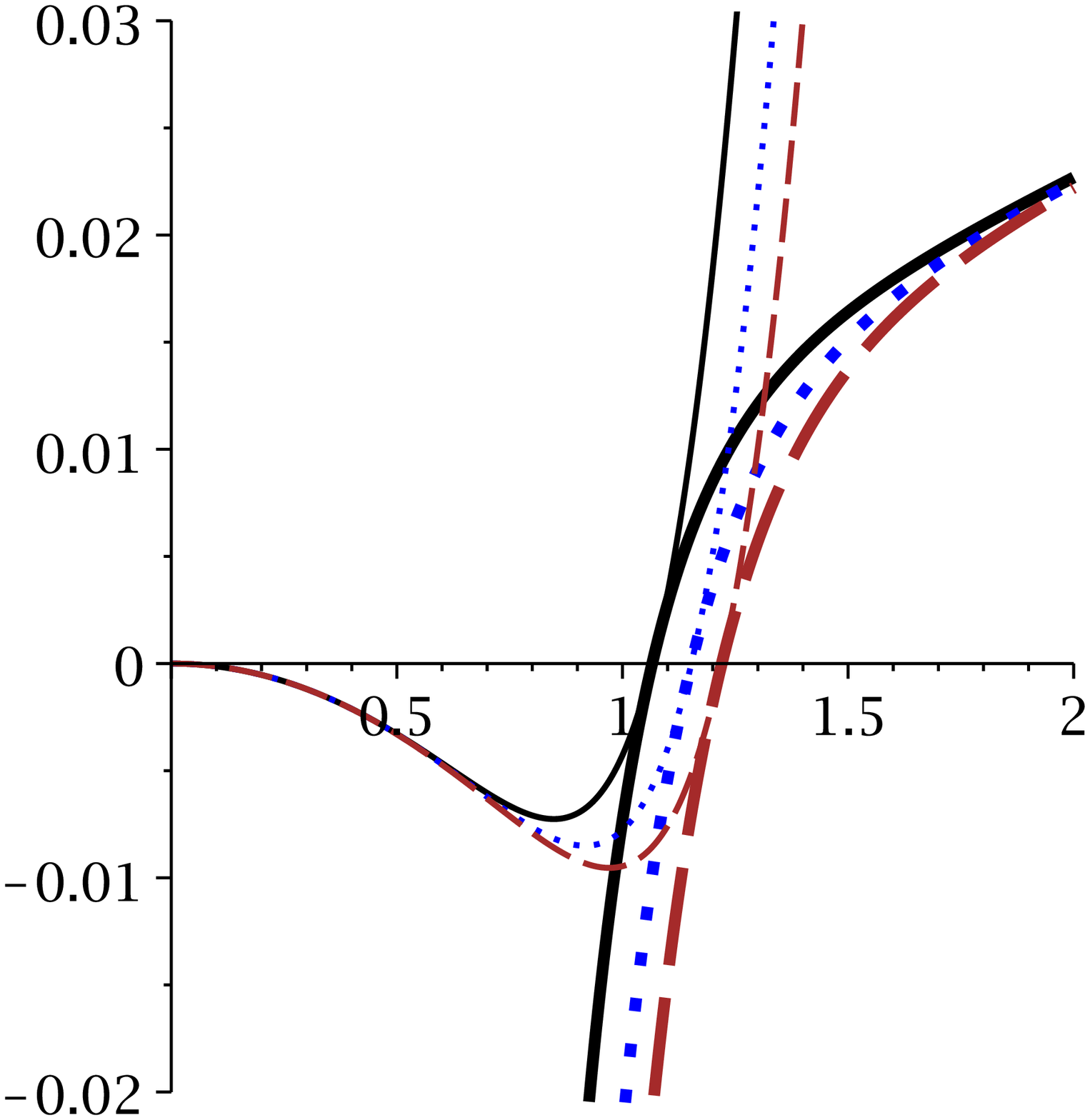}%
\end{array}
$%
\caption{\textbf{PMI branch:} $C_{Q}$ and $T$ (bold lines) versus $r_{+}$
for $k=1$, $l=1$, $\protect\varepsilon=0.2$ and $s=0.7$. \newline
left panel: $q=1$, $\protect\xi=0.5$ (continues line), $\protect\xi=1$
(dotted line) and $\protect\xi=5$ (dashed line). \newline
right panel: $\protect\xi=1$, $q=0.9$ (continues line), $q=1.3$ (dotted
line) and $q=1.7$ (dashed line).}
\label{Fig4}
\end{figure}

\begin{figure}[tbp]
$%
\begin{array}{ccc}
\epsfxsize=5.5cm \epsffile{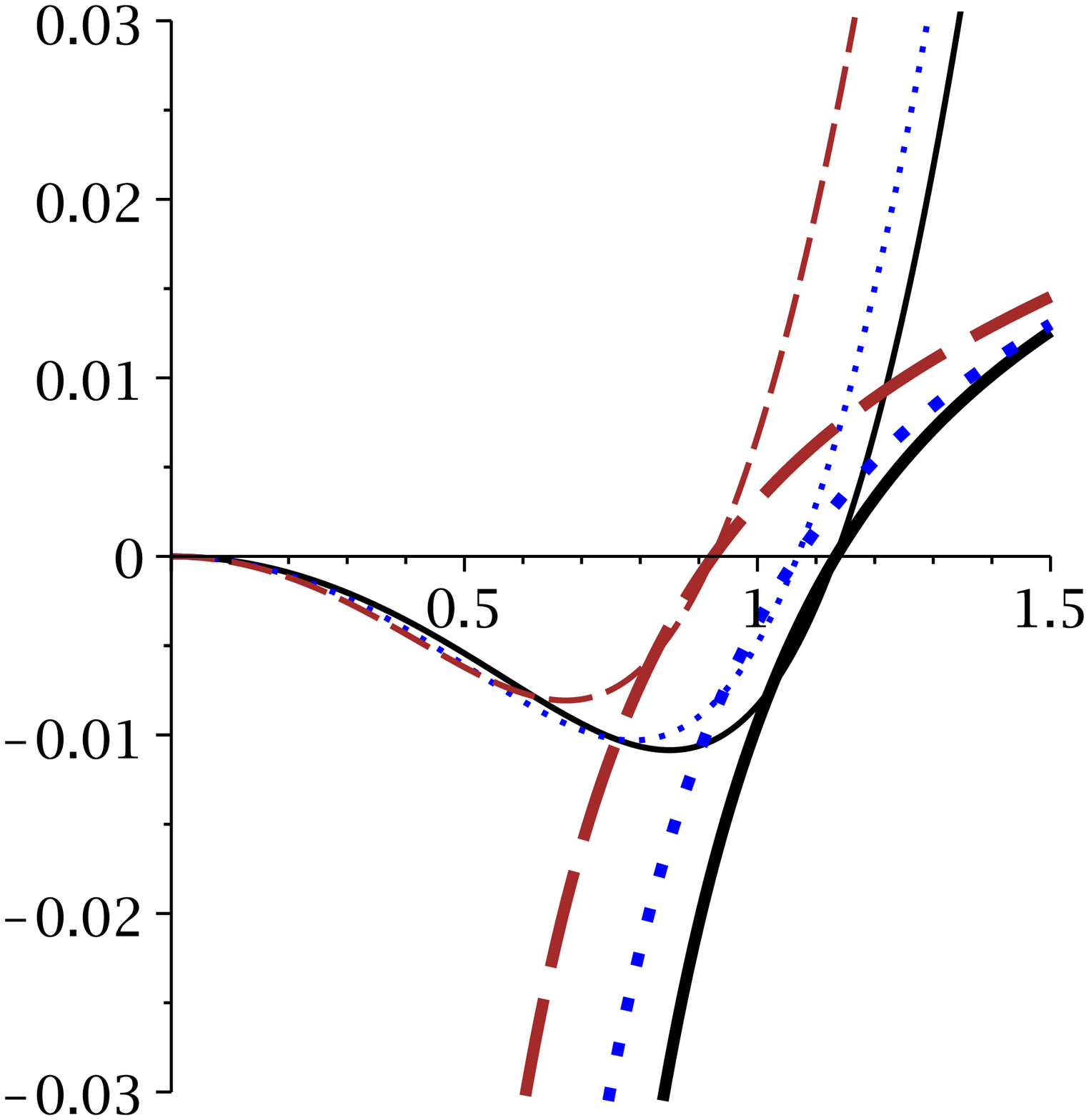} & \epsfxsize=5.5cm %
\epsffile{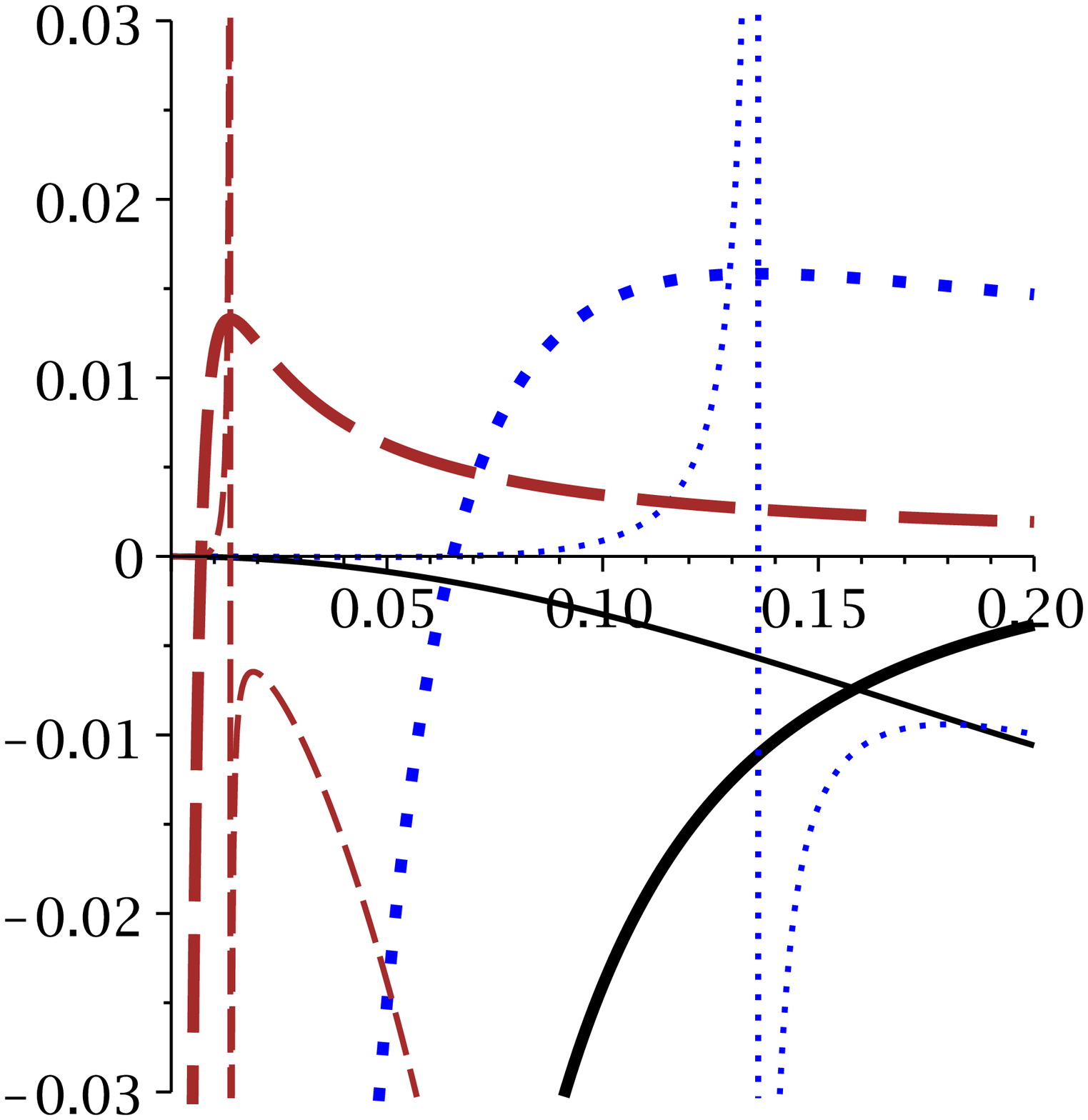} & \epsfxsize=5.5cm \epsffile{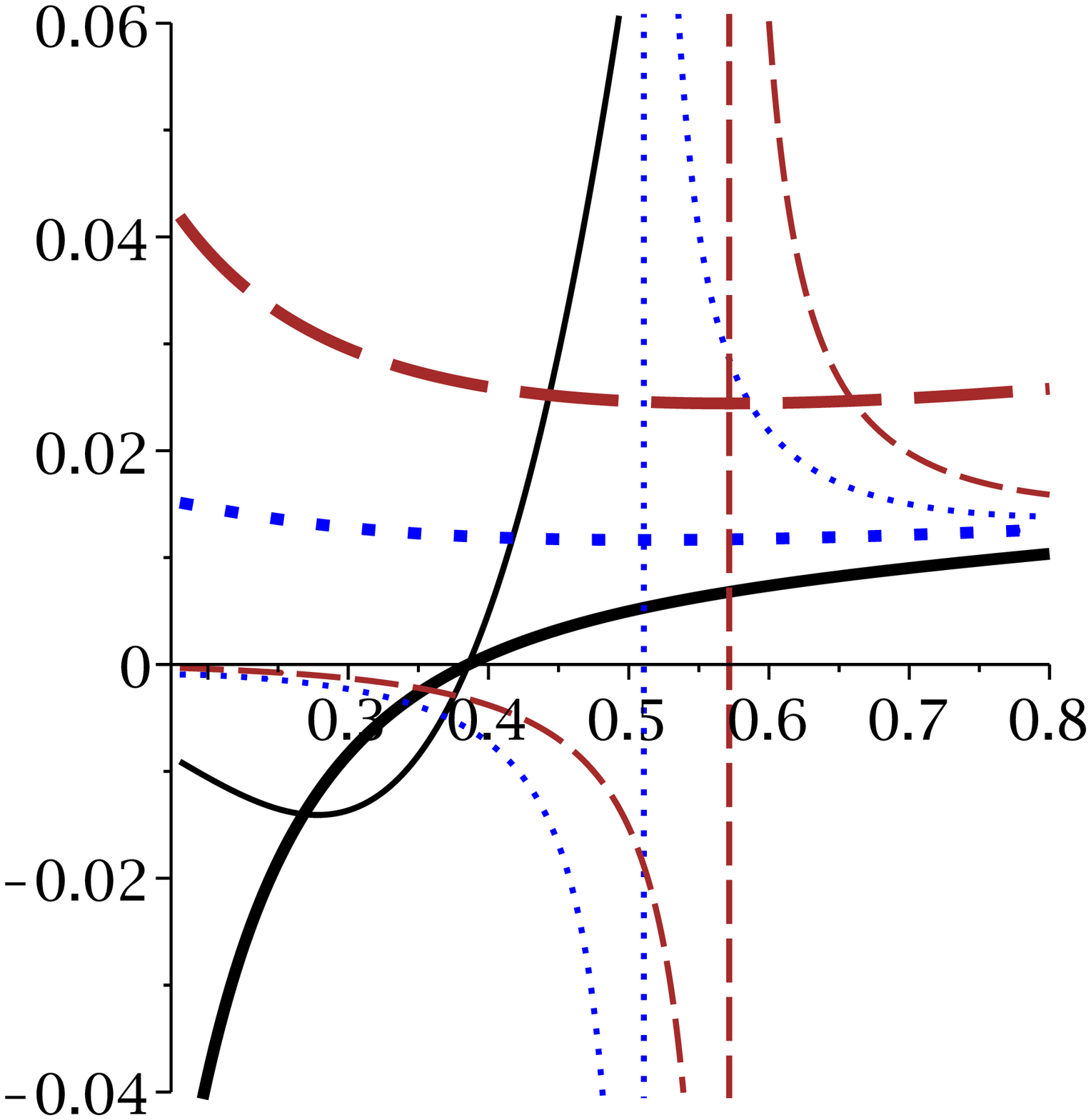}%
\end{array}
$%
\caption{\textbf{PMI branch:} $C_{Q}$ and $T$ (bold lines) versus $r_{+}$
for $k=1$, $l=1$, $\protect\varepsilon=0.2$, $q=1$ and $\protect\xi=1$ "for
different scales". \newline
left panel: $s=0.9$ (continues line), $s=1$ (dotted line) and $s=1.1$
(dashed line). \newline
middle and right panels: $s=1.3$ (continues line), $s=1.4$ (dotted line) and
$s=1.45$ (dashed line).}
\label{Fig5}
\end{figure}

In order to study the thermodynamical behavior of the system, we use Eq. (%
\ref{loop}), in which $f(\varepsilon)$ has an exponential form. Considering
these two rainbow functions, we plot following diagrams to study the effects
of variation of different parameters on stability conditions and phase
transition of the obtained solutions (Figs. \ref{Fig1}-\ref{Fig5}).

%%%%%%%%%%%%%%%%%%%%%%%%%%%%%%%%%%%%%%%%%%%%%%%%%%%%%%%%%%%%%%%

%%%%%%%%%%%%%%%%%%%%%%%%%%%%%%%%%%%%%%%%%%%%%%%%%%%%%%%%%%%%%%%

%%%%%%%%%%%%%%%%%%%%%%%%%%%%%%%%%%%%%%%%%%%%%%%%%%%%%%%%%%%%%%%

%%%%%%%%%%%%%%%%%%%%%%%%%%%%%%%%%%%%%%%%%%%%%%%%%%%%%%%%%%%%%%%

%%%%%%%%%%%%%%%%%%%%%%%%%%%%%%%%%%%%%%%%%%%%%%%%%%%%%%%%%%%%%%%

%%%%%%%%%%%%%%%%%%%%%%%%%%%%%%%%%%%%%%%%%%%%%%%%%%%%%%%%%%%%%%%

%%%%%%%%%%%%%%%%%%%%%%%%%%%%%%%%%%%%%%%%%%%%%%%%%%%%%%%%%%%%%%%

%%%%%%%%%%%%%%%%%%%%%%%%%%%%%%%%%%%%%%%%%%%%%%%%%%%%%%%%%%%%%%%

%%%%%%%%%%%%%%%%%%%%%%%%%%%%%%%%%%%%%%%%%%%%%%%%%%%%%%%%%%%%%%%

%%%%%%%%%%%%%%%%%%%%%%%%%%%%%%%%%%%%%%%%%%%%%%%%%%%%%%%%%%%%%%%

In case of BI type models, for specific values of different parameters,
there is a region in which the temperature and heat capacity are negative.
Black hole solutions are not physical in this region. In this case, the heat
capacity enjoys a root, $r_{+0}$, in which for $r_{+}>r_{+0}$ , BI type
black holes are in stable state with positive temperature. $r_{+0}$ is a
decreasing function of $\xi$ (Fig. \ref{Fig1}) and an increasing function of
electric charge (Fig. \ref{Fig2}). Interestingly, for small values of $q$,
heat capacity may enjoy a divergency which indicates a second order phase
transition. Here, a phase transition of smaller to larger black holes takes
place (left panel of Fig. \ref{Fig2}). Surprisingly, the plotted diagram for
temperature in this case, shows the existence of subcritical isobar. The
presence of subcritical isobars is observed for Van der Waals like
liquid/gass systems. Such behavior for black holes is only observed in cases
of considering cosmological constant as thermodynamical pressure \cite%
{extended}. Here, without the use of analogy between cosmological constant
and thermodynamical pressure, we found the properties of critical point.

In addition, we see that for suitable choices of different parameters and
small values of nonlinearity parameter, there are one root and two extrema
in temperature diagram; one minimum and one maximum. These extrema present
themselves as divergencies in heat capacity diagrams. The stable states
exist between the root and smaller divergency and after the larger
divergence point (Fig. \ref{Fig3}). Between two divergencies, the heat
capacity is negative while the temperature is positive, and therefore, in
this region unstable state exists. This instability switches to stable state
as heat capacity meets the divergencies. In other words, a phase transition
of smaller unstable to larger stable solutions takes place at the larger
divergence point, while a phase transition of larger unstable to smaller
stable occurs at the smaller divergence point. Increasing the nonlinearity
parameter leads to vanishing of these phase transition points (Fig. \ref%
{Fig3}).

It is worthwhile to mention that the small values of nonlinearity
parameter represent strong nonlinearity for the system. Therefore,
as the nonlinearity of the system increases (nonlinearity
parameter decreases), the thermodynamical structure of the
solutions will be modified. In addition, numerical evaluation
shows that the root of heat capacity in LNED theory is bigger
comparing to other theories of BI family. This indicates that
physical stable solutions are obtained in higher values of horizon
radius for this theory of NED comparing to BINED and ENED. In
opposite, the divergence point in LNED is located at smaller
horizon radius which shows that black hole solutions in presence
of LNED acquire thermal stability faster comparing to other two BI
models.

For PMI case, for specific values of different parameters, like BI case, a
root is observed which is a limitation bound between non-physical and
physical states. The value of root is a decreasing function of $\xi$ (Fig. %
\ref{Fig4} left panel) and an increasing function of electric charge (Fig. %
\ref{Fig4} right panel). Interestingly, regarding the variation of $s$, for $%
0.5<s<s_{c}$ ($s_{c}\approx 1.3$ for considered parameters), only a root
(mentioned bound point) exists (Fig. \ref{Fig5} left panel). As for $s=s_{c}$%
, there are two extrema, one minimum and one maximum, and also one root for
the temperature (Fig. \ref{Fig5} middle and right panels). Between root and
smaller divergence point and after larger divergence point, heat capacity is
positive and system is in stable state, whereas between two divergencies
system has negative heat capacity (with positive $T$), hence, black holes
are not stable. Another interesting effect is that for $s_{c}<s<1.5$, the
smaller divergence point is a decreasing function of $s$ while larger
divergency is an increasing function of it (Fig. \ref{Fig5}).

\section{Closing Remarks}

In this paper, we have considered gravity's rainbow in presence of various
models of NED. At first, $4$-dimensional black hole solutions for these
configurations were derived, and then, related conserved and thermodynamic
quantities were calculated. It was shown that some of the conserved and
thermodynamical quantities were modified due to the contribution of
gravity's rainbow. Despite these modifications, the first law of
thermodynamics was valid for these black hole solutions.

Next, we have studied the stability of the solutions and phase transition
points in context of canonical ensemble. The employed nonlinear
electromagnetic fields in this paper were categorized into two types: BI
type includes Born-Infeld, logarithmic and exponential forms and PMI model
which is power law generalization of Maxwell Lagrangian.

In case of BI types, we have found a lower bound for the horizon radius, $%
r_{+0}$, in which the black holes are not physical for
$r_{+}<r_{+0} $. Interestingly, for suitable choices of different
parameters, we have found a second order phase transition point
which had characteristics of $T-V$ diagrams for critical pressure
(subcritical isobar was observed). Then, by employing different
parameters, we have found two extrema and a root in temperature,
and one root and two divergencies in heat capacity diagrams. In
this case, there existed two phase transitions of smaller unstable
to larger stable and larger unstable to smaller stable. The phase
transitions took place at divergencies of the heat capacity. It
was pointed out that the largest root and the smallest divergence
point of the heat capacity belonged to the logarithmic form of
NED.

For PMI case, similarly, stable physical and unstable non-physical states
were observed for a range of $s$. Interestingly, for another range of this
parameter the thermodynamical behavior was modified. A root, a maximum and a
minimum were observed for the temperature. In places of these extrema (heat
capacity enjoyed the existence of divergencies), two phase transitions of
medium unstable to smaller or larger stable black holes took place. Another
interesting property of this matter-field was the effects of variation of
the $s$ on divergencies of the heat capacity. For specific range of $s$, the
smaller divergence point was a decreasing function of $s$ while the larger
divergency was an increasing function of nonlinearity parameter.

It is evident that BI types and PMI models have different effects and
contributions to thermodynamical behavior of the black hole system. In other
words, considering these two classes of NED leads to different modifications
and properties for the system. In case of BI family of NED, the theory under
consideration will indicate the place of formation of the stable solutions.

It will be worthwhile to study obtained solutions in this paper in context
of extended phase space and investigate both modifications of gravity's
rainbow and nonlinear electromagnetic field on critical behavior of the
system. In addition, generalization to higher dimensions is another
interesting work. We left these issues for the forthcoming work.

\begin{acknowledgements}
We would like to thank the anonymous referee for useful
suggestions and enlightening comments. We also thank Shiraz
University Research Council. This work has been supported
financially by the Research Institute for Astronomy and
Astrophysics of Maragha, Iran.
\end{acknowledgements}

\end{document}